\documentclass[sigconf,authorversion]{acmart}
\usepackage[english]{babel}
\usepackage[ansinew]{inputenc}

\AtBeginDocument{%
  \providecommand\BibTeX{{%
    \normalfont B\kern-0.5em{\scshape i\kern-0.25em b}\kern-0.8em\TeX}}}

\setcopyright{acmcopyright}
\copyrightyear{2020}
\acmYear{2020}
\setcopyright{acmlicensed}\acmConference[ARES 2020]{The 15th International Conference on Availability, Reliability and Security}{August 25--28, 2020}{Virtual Event, Ireland}
\acmBooktitle{The 15th International Conference on Availability, Reliability and Security (ARES 2020), August 25--28, 2020, Virtual Event, Ireland}
\acmPrice{15.00}
\acmDOI{10.1145/3407023.3407026}
\acmISBN{978-1-4503-8833-7/20/08}

\begin{document}

\title{An Overview of Limitations and Approaches in Identity Management
}

\author{Daniela P{\"o}hn}
\email{daniela.poehn@unibw.de}
\author{Wolfgang Hommel}
\email{wolfgang.hommel@unibw.de}
\affiliation{%
  \institution{Universit{\"a}t der Bundeswehr M{\"u}nchen, Research Institute CODE}
  \city{Munich}
 \country{Germany}
}

\renewcommand{\shortauthors}{P{\"o}hn and Hommel}

\begin{abstract}
Identity and access management (I\&AM) is the umbrella term for managing users and their permissions. It is required for users to access different services. These services can either be provided from their home organization, like a company or university, or from external service providers, e.\,g., cooperation partners. I\&AM provides the management of identifiers with the attributes, credentials, roles, and permissions the user has. Today, the requirements have evolved from simply accessing individual web services in the internet or at a company to the majority of all IT services from different service providers with various accounts. Several identity management models have been created with different approaches within.

In order to adjust to heterogeneous environments, use cases, and the evolution of identity management, this paper extends known requirements for identity management. Existing models and approaches for identity management are mapped to the derived requirements. Based on the mapping, advantages, disadvantages, and gaps are identified. Current approaches suffer, as an example, from trustworthiness and liability issues. Interoperability issues are even more inherent as the approaches partly develop apart, forming an heterogeneous environment. The results from this analysis emphasize the need for one holistic identity management framework.
\end{abstract}



\maketitle

\section{Introduction}\label{introduction}

Medium-sized and large organizations, such as universities and companies, typically provide several information and communications technology services to their members. Usually, a username as identifier is assigned to each member. All services, e.\,g., email, file, and web collaboration, can then be used by the combination of username and some sort of credential. In the backend, username, salted and hashed password, and further information, called attributes, are stored centrally. Organizations either use lightweight directory access protocol (LDAP) or other database management systems for this. Even though a local Identity \& Access Management (I\&AM) seems trivial, it gets challenging, if interfaces to other entities appear, e.g., with collaborations or joint ventures.

When members of an organization want to access external services, like collaboration platforms, either their user accounts are doubled or Federated Identity Management (FIM) is used. FIM is an arrangement between multiple entities, in order to let users use the same identification data as in their home organization to obtain access to the services of the provided entities within the trust boundaries. FIM is mostly based on the protocol Security Assertion Markup Language (SAML)~\cite{samloverview}, which is common in the R\&E field as well as in national federations, or on OpenID Connect (OIDC)~\cite{openidconnect}. OIDC is the authentication layer on top of the authorization protocol OAuth~\cite{RFC6749}, while SAML incorporates both aspects. OAuth and OIDC are widely used in today's web services, e.\,g., Google Application Interfaces (APIs). Each user has at least some kind of home organization, called Identity Provider (IdP) in SAML or Relying Party (RP) in OIDC. The user wants to access a service from a Service Provider (SP), respectively OpenID Provider (OP). The set of IdPs and SPs that collaborate for a specific reason is commonly referred to as federation. Federations are especially customary in SAML~\cite{samloverview}, however an OIDC specification is currently developed~\cite{oidcfederation}. While federations in industry are relatively small, many national research and education networks (NRENs) operate large authentication and authorization infrastructures (AAIs) with hundreds of organizations. Technical trust is established by pre-sharing metadata, i.\,e., information about the entity in extensible markup language (XML) files. Another form of trust is set into contracts, e.\,g. between federation operator and participating entity.

Due to limitations of the rather static SAML federations, inter-federations, like eduGAIN~\cite{edugainstatus}, were established and dynamic concepts, e.\,g. interfederation metadata exchange~\cite{metadataexchangeconcepts} and TrustBroker~\cite{infocomp2014}, were developed. As industry tends to use OIDC, organizations set up OIDC in parallel or even changed completely. At the same time eID federations with SAML were established~\cite{stork}. While SAML federations have trust established in out-of-bounds mechanisms, other mechanisms need to be set up with OIDC. This led to the growth of different assurance frameworks and registries~\cite{iana}, while, at the same time, a Domain Name System (DNS) like structure for discovery of the SP and verification of trust scheme membership, i.\,e. LIGHTest~\cite{mci/Rossnagel2017}, was developed.

I\&AM as well as FIM focus on organizations. Since some users are at least concerned about their identity and privacy, user-centric solutions were developed. A privacy policy of a service provider summarizes the usage of collected personal user data, which can now be gained by EU general data protection regulation (GDPR). At that time, most sites described user information in little understandable way. User-centric identity management (UCIM) targeted these shortcomings by a clear presentation and intuitive evaluation of privacy policies. One first application was Microsoft Passport, an IdP in a Microsoft environment, which could be used as universal IdP in the internet. As Microsoft was integrated in every communication, trust and privacy were drawbacks~\cite{1212687}. As a next step, Windows Cardspace was developed, which could make use of other IdPs with SAML. The user had several infocards, either created by himself or managed by an IdP. The application possessed several security bugs though~\cite{4299788}. User Managed Access (UMA)~\cite{uma}, an OAuth-based standard, even goes one step further. It enables the user to control the authorization of data sharing and other protected resources. Sovrin~\cite{sovrin} sees Self-Sovereign Identities as the next step of evolution. The user must be the ultimate owner of a self-sovereign identity. Therefore, a self-sovereign identity is fully autonomous in management as well as administration, while the user is the single source of truth. This paradigm is realized by decentralized domains; inherent this means the use of distributed ledger technologies (DLTs), i.\,e., blockchain~\cite{8776589}.

Hence, orthogonal approaches are currently developed and applied. SAML is rather static with solid trust boundaries, in contrast to OIDC. DNS-based structures facilitate stronger trust, based on certificates and eIDs. While UMA can be applied to OIDC, self-sovereign identities (SSIs) can be used for different use cases. The bouquet of identity management approaches lead to different problems. Administrators need to take care of different solutions in organizations. Due to this, getting an overview and applying similar policies gets cumbersome. Also, if companies join each other or establish projects, it might be the case, that their identity management solutions are incompatible. This would require yet another system to manage. In addition, not all requirements from the organizations might been fulfilled by the systems, leading to further tools. Besides administrators, also end-users and other stakeholders have lots of solutions to decide from. The end user might end up using several different systems to manage his identities. Therefore, it is important to have an overview, showing advantages and disadvantages, as well as gaps. This overview can help the stakeholders to choose the best fitting approach and identify needed tools.

In order provide this overview, we extend known requirements for identity management in Section~\ref{sec:motivation}. The related work is divided into the established models~\cite{5689468} centralized identity management, FIM, and UCIM in Sections~\ref{sec:sota} to~\ref{sec:ucim}. The existing approaches are analysed based on the requirements and gaps are identified. The paper is concluded by an outlook and summary of the analysis.

\section{Requirements of Identity Management}\label{sec:motivation}

In order to understand the usage of the different identity management approaches and discover possible gaps, various requirements are identified. According to Torres et al.~\cite{6275425}, the main requirements are Usability, Interoperability, Functionality, Trustworthiness, Security, Mobility, Privacy, Law Enforcement, and Affordability. Boujezza et al.~\cite{7507266} refer to these requirements as bases for IoT. Ferdous and Poet leaf out Mobility~\cite{6266958}, while Gao et al.~\cite{7423325} concentrate on the core functionalities.  The authors of~\cite{8742375} see User control and consent, Minimal disclosure for a constrained use, Justifiable parties, Directed Identities, Design for a pluralism of operators and technology, Human integration, and Consistent experience across contexts as requirements for SSI.

Regarding SSIs, the management of users has to be another requirement. Besides management of identities, other functionalities might be needed for today's identity management models, approaches, and protocols to be interoperable. Therefore, the requirement Functionality is divided into Management and Functionality. The location of the user data is renamed to Portability. With the GDPR in Europe, the user can ask to get his account data exported and then imported to another platform. Therefore, the user data needs to be portable from one system to another. Additionally, the user should be able to make use of different devices, see requirement Design for a pluralism of operators and technology. Affordability is split into Integration into the existing system and Scalability. As more and more devices and thereby identities are created, scalability becomes increasingly important. 

In consequence, the following requirements are used in further sections to evaluate the different approaches. These requirements extend the requirements gathered from  Torres et al.~\cite{6275425}. They are already condensed and comprise different sub-requirements, as shown in Table~\ref{tab:req}.

\begin{table}
\caption{Requirements for Identity Management}
\label{tab:req}
\begin{tabular}{llp{4.3cm}}
No. & Name & Description  \\ \hline
REQ1 & Management & of Identities, Automation \\
REQ2 & Usability & User Interface, Reduced Complexity, Consistent Experience \\
REQ3 & Interoperability & Interoperability between Protocols, Models, Silos \\
REQ4 & Scalability & Mobility of Environment \\
REQ5 & Functionality & Translation Services, Proxies, Bridges, Group Management \\
REQ6 & Trustworthiness & Trust Management, Segregation of Power, Policies \\
REQ7 & Security &  Basic and Multi-Lateral \\
REQ8 & Portability & of Accounts and Systems \\
REQ9 & Privacy & Anonymity, Pseudonymity, Transparency, Controlability, Consent, Data Minimization \\
REQ10 & Liability & Law-enforcement, Digital Evidence, Data Retention \\
REQ11 & Integration & Affordability, Integration into Processes and Existing Infrastructure \\
\end{tabular}
\end{table}

\section{Centralized Identity Management}\label{sec:sota}

As identity management advances in different directions, selected areas of related work are covered in the following. In order to structure the work, we use different identity management models:
\begin{itemize}
\item centralized / network-centric,
\item federated / application-centric,
\item decentralized / user-centric.
\end{itemize}
These models have further characteristics, used in different use cases with different protocols. Even though centralized identity management was the first evolutionary step after single instances, it is still used. Mostly companies, which control all their own identities, use single sign-on between services. If they outsourced IdM, it is most likely FIM. Also, companies with different locations may have one federation. Within a centralized identity management, the company is in control of law-enforcement [REQ10] and there is no need for further interoperability [REQ3]. If cooperations are established, it looks differently.

\subsection{Centralized Identity Management with DNS}

The DNS is the hierarchical and decentralized naming system for computers, services, and other resources connected to a network. Most commonly used, it translates memorable domains to numeric internet protocol (IP) addresses needed for locating and identifying computer services and devices with the underlying network protocols. In an identity federation, the home organization is discovered by specific protocols and partly also services, i.\,e., a trusted third party having a list of possible entities in form of metadata. Furthermore, in order to make use of different protocols at one entity, at least bridges need to be established. A way to improve the situation could be the usage of DNS. The EU project LIGHTest~\cite{Rossnagel2019} \cite{mci/Rossnagel2017} is based on DNS and DNS Secure Extensions (DNSSEC) as a method of discovery and trust. Therefore, a trust scheme publication authority (TSPA)~\cite{mci/Wagner2019} needs to be set up. The approach can make use of role-based access control (RBAC) as a form of delegation, while it harmonizes the delegation data format~\cite{mci/Wagner2017}. As drawbacks, eIDAS or some sort of certificates are needed, which also means fixed Level of Assurance (LoA), therefore [REQ6] is only partly fulfilled, and a centralized structure, similar to SAML. Even though it can be used with internet of things (IoT)~\cite{mci/Jeyakumar2019}, [REQ3] is not fulfilled.

\subsection{Centralized Identity Management with DNS for IoT}

Wang and Wen~\cite{6192955} present a privacy-enhanced DNS-scheme for IoT.  DNS was not designed with all the security complications of today's world in mind. Therefore, the authors introduce new features. Nevertheless, the approach does not take caching into account, which contradicts [REQ7] and [REQ9]. Additionally, fresh domain names are needed after several steps [REQ1]. The approach of Lee et al.~\cite{7423411} describes the auto-confi\-gu\-ra\-tion of IoT devices with the device's category and model in IPv6 environment, as the neighbor discovery is used. It is, in contrast, inefficient due to manual configuration of the DNS names for IoT devices [REQ1]. Though the discovery via DNS is widely known as well as possible attacks, no authorization is provided. The size of messages can get a problem as well as the scalability [REQ4], if too many features are added to DNS. Therefore, changes are difficult [REQ11]. According to the presented approaches, identity management with DNS can be used in use cases, where no anonymity is needed, but where eIDAS, fixed trust values, and a centralized structure for certificates is in place.

\subsection{Centralized Identity Management for IoT}

Not only humans need to be identified, but also things, like computers and IoT devices. IoT devices often have limitations, like bandwidth or central processing unit (CPU). Therefore, different protocols are used, as shown in~\cite{7968561}. Belran and Skarmeta~\cite{7845482} give an overview of the IETF protocol Authentication and Authorization for Constrained Environments (ACE)~\cite{RFC7744}, which is extended by several Internet-Drafts (I-Ds) [REQ3]. The authors of~\cite{7185289} identify and authenticate IoT objects in a natural context using DNS with time property, name, and location resolution. Efficiency degradation is avoided, but the life cycle is not taken into account [REQ11]. P. N. Mahalle's thesis~\cite{04edd2148d064f3fa0e2addc0d6da508} is about identity management for IoT devices. Even though the context of the identifier is specified, users and their relationship to objects are not mentioned [REQ1] [REQ11].  \cite{10.1145/2490428.2490456} focuses on cloud-based IoT. While it takes the life cycle of IoT things into account, it does not consider the owner [REQ11].

\subsection{Conclusion from Centralized Identity Management}

The results are summarized in Table~\ref{tab:iot}. Many approaches do not fulfill the requirement Interoperability [REQ3]. This is comprehensible, as they are solely centralized. Additionally, centralized IdM approaches seem to lack Trustworthiness [REQ6], Management [REQ1], and Integration [REQ11] in several cases. Especially Trustworthiness and Integration come inherent with the model. Portability [REQ8] is difficult to estimate, as the approaches are mostly described for computer use cases.

\begin{table}
\caption{Overview of Approaches for Centralized Identity Management}
\label{tab:iot}
\begin{tabular}{p{1.2cm}p{2.5cm}p{3.5cm}}
Ref & Name & Requirements not fulfilled  \\ \hline
 & Centralized IdM & [REQ3] \\
\cite{Rossnagel2019} \cite{mci/Rossnagel2017} & LIGHTest & [REQ6] [REQ3] \\
\cite{mci/Wagner2019} & TSPA & [REQ6] [REQ3] \\
\cite{mci/Wagner2017} & Delegation & [REQ6] [REQ3] \\
\cite{mci/Jeyakumar2019} & LIGHTest for IoT & [REQ6] [REQ3] \\
\cite{6192955} & Privacy-enhanced DNS-scheme & [REQ1] [REQ7] [REQ9] \\
\cite{7423411} & Autoconfiguration of IoT devices & [REQ1] \\
\cite{RFC7744} & ACE & [REQ3] \\
\cite{7185289} & Natural context & [REQ11] \\
\cite{04edd2148d064f3fa0e2addc0d6da508} & PhD thesis & [REQ1] [REQ11] \\
\cite{10.1145/2490428.2490456} & Cloud-based IoT & [REQ11] \\
\end{tabular}
\end{table}

\section{Federated Identity Management}

Federated identity management is an arrangement between multiple entities within certain trust boundaries. It lets users use the same identification as in their home organization to obtain access to the services of the provided entities. In many constellations, a trusted third party is needed. Especially SAML federations tend to have scalability problems [REQ4]. Between different FIM approaches, interoperability is a problem [REQ3]. Liability [REQ10] depends on all entities, while the trusted third party could provide additional functionality, though it is not the case with the following approaches [REQ5]. Bargh et al.~\cite{MortazaS.Bargh2011} show that the growing number of IdPs and SPs is a well-known problem of SAML in academic world. Different approaches, therefore, try to improve the scalability of trust and metadata exchange as well as trust negotiation in federated identity management.  IdMRep by Arias~Cabarcos et al.~\cite{Arias:2013:IdMRep} introduces dynamic trust values. The presented approach does not work with new members [REQ6]. Additionally, the accumulated amount of data at the trust engine can become to a single point of failure in large-scale infrastructures [REQ4]. The approach Dynamic Identity Federations by Md. Sadek Ferdous and Ron Poet~\cite{dynfed} focuses on fully automated establishment of federations and trust. Though the technical establishment is fully automated, the user has to enter the EntityID of the IdP and needs to generate a code, which is not user-friendly [REQ2]. Furthermore, the division into trusted, semi-trusted, and untrusted is rather coarse-grained [REQ6], and a further database per entity is needed [REQ11]. Another approach is called TrustBroker by P{\"o}hn et al.~\cite{ifipsec2014}, which is based on SAML, though the generic concepts can be applied to FIM [REQ3]. There are also other federation approaches. OLYMPUS by Moreno et al.~\cite{8766357} is privacy preserving, while it allows distributed IdPs. Nevertheless, it misses [REQ3]. Kang and Khashnobish~\cite{5070648} approach for a peer-to-peer (P2P) federation is decentralized, but has scalability issues in large-scale infrastructures with many entities and federation servers [REQ4].  Dhungana et al.~\cite{6544386} provide an IDM framework for cloud networking based on UMA with OIDC. It thereby allows federated identity management without the standard of OIDC federations~\cite{oidcfederation} [REQ11].
UnlimitID by Isaakidis et al.~\cite{Isaakidis:2016:UPF:2994620.2994637} utilizes attribute-based anonymous credentials, which are based on algebraic Message Authentication Codes (MACs). The approach is usable with OAuth and let pseudonyms expire. Though for this short time users might be linkable [REQ9]. Furthermore, in some use cases pseudonyms might not be allowed [REQ11]. The usage of MACs seems like a step back. These approaches show that certain aspects are regarded, which can be solved by federated identity management, but not all. Additionally, especially FIM with SAML is difficult to port to mobile applications and IoT devices [REQ8].

\subsection{Federated Identity Management with DNS}

ID4me~\cite{id4me} combines DNS with OIDC. It utilizes the protocol Automated Certificate Management Environment (ACME), which automatically verifies the ownership of a domain and also simplifies the issuance of certificates. The DNS challenge is used as discovery method, secured with DNSSEC. The communication is secured by DNS-based Authentication of Named Entities (DANE), which is only rarely used in practice. ID4me has the roles identity agent, which administrates the attributes and is identified by DNS, and identity authority, which is used for authentication. As a consequence, the user has to trust the identity agent [REQ6]. The idea to use DNS is not new. Hulsebosch et al.~\cite{1690550} present the combination for effective and trusted user controlled light-path establishment in grid environment. Thereby, DNS tree is used to establish technical trust between IdP and Optical Network Service Provider (ONSP) dynamically within a virtual organization (VO), a kind of federation [REQ3].

\subsection{Federated Identity Management with eID}

The EU Regulation 910/2014, also known as eIDAS, is effective since September 2018. The authors of~\cite{8543142} show the federated identity architecture with eIDs. The approach of Berbecaru et al.~\cite{7912677} describes the connect between eIDAS and STORK. STORK~\cite{5514447} \cite{Ribeiro:2018:SRH:3280301.3280330} is the eID federation based on eID. It features the European electronic identity interoperability platform, which relies on SAML with one node per country. Thereby, it is heterogeneous and a large-scale infrastructure. Several approaches describe the integration of academic attributes in the eIDAS infrastructure without regarding the inter-federation eduGAIN, which is the de-facto standard federation in the academic world~\cite{8540674} \cite{info10060210} [REQ11]. Alonso et al.~\cite{8754671} proposed an identity model to connect FIWARE services authenticated with OAuth 2.0 to eIDAS nodes and thereby meet the regulations. It nevertheless is the solution for one specific use case [REQ3].

\subsection{Federated Identity Management as a Service}

Other approaches focus on federation as a service, which utilizes cloud infrastructure. The framework AIDF by Zouari and Hamdi~\cite{7746120} makes use of a trust broker, like in~\cite{ifipsec2014}, in cloud. IdP and SP register at a central party. The SP can tell, which IDs and attributes he wants and he can at the same time subscribe to claim transformation and attribute mapping service. Attribute mapping is necessary, if different semantic or/and syntax of attributes is used. A simple example is date format. The central service automatically establishes a link between IdP and SP after consent. Some sort of trust assurance is established by National Institute of Standards and Technology (NIST) LoA, which is currently published in the third revision of the special publication 800-63 from 2017. As the LoA is revised, the framework needs updates as well. Furthermore, not only the NIST LoA could be applied, but also other standards [REQ6]. Additionally, there exists no trust assurance for IdPs, solely for SPs. The same drawback with NIST LoA has the approach of Zefferer et al.~\cite{8356430}, which links identities to eIDs. Even though the federations have a strong legal foundation as eIDs are used, it is highly customized for this one use case [REQ3].

\subsection{Conclusion from Federated Identity Management}

A summary of FIM approaches is found in Table~\ref{tab:fim}. FIM itself lacks Interoperability [REQ3], Functionality [REQ5], and Liability [REQ10]. Liability often comes with policies and other agreements. Interoperability can be archived with additional tools, but the protocols as they are act like silos. Looking at the protocol SAML, Scalability [REQ4] and Portability [REQ8] are also missing. The approaches described in this paper often have drawbacks in Interoperability [REQ3] and Trustworthiness [REQ6]. Also, Integration [REQ11] is often stated, while it is partly difficult to estimate Liability [REQ10] as it depends on the implementation and established processes. As the users access services from other organizations, trust needs to be established between home organization and service provider, user and service provider, and user and home organization. If additional entities are involved, they need to be involved in the trust establishment as well.

\begin{table}
\caption{Overview of Approaches for Federated Identity Management}
\label{tab:fim}
\begin{tabular}{p{1.2cm}p{2.5cm}p{3.5cm}}
Ref & Name & Requirements not fulfilled  \\ \hline
 & FIM & [REQ3] [REQ5] [REQ10] \\
\cite{Arias:2013:IdMRep} & IdMRep & [REQ4] [REQ6] \\
\cite{dynfed} & Dynamic Identity Federations & [REQ2] [REQ6] [REQ11] \\
\cite{ifipsec2014} & TrustBroker & [REQ3] \\
\cite{8766357} & OLYMPUS & [REQ3] \\
\cite{5070648} & P2P federation & [REQ4] \\
\cite{6544386} & IDM framework & [REQ11] \\
\cite{Isaakidis:2016:UPF:2994620.2994637} & UnlimitID & [REQ9] [REQ11] \\
\cite{id4me}  & ID4me & [REQ6] \\
\cite{1690550} & VO & [REQ3] \\
\cite{8540674} \cite{info10060210}  & Academic eIDAS & [REQ11] \\
\cite{8754671} & FIWARE eIDAS & [REQ3] \\
\cite{7746120} & AIDF & [REQ6] \\
\cite{8356430} & eID federations & [REQ3] [REQ6]\\
\end{tabular}
\end{table}

\section{User-Centric Identity Management}\label{sec:ucim}

User-Centric Identity Management puts the user in control of their own identities. It has evolved from simple clients on a PC to UMA and SSI, which are described in the following. Although it gives more power to the user and thereby adds value to [REQ9], the interoperability with other approaches is low [REQ3], as [REQ5] is missing. [REQ10] depends on the client and the provider.

\subsection{User-Centric Identity Management for IoT}

Do van Thuan et al.~\cite{7021724} present a user centric identity management for IoT with four identifiers, i.\,e., idpID, domIDPart, devIDPart, and userIDPart. They describe different relationships as well, e.\,g., serving people in general, serving one person permanently, serving multiple persons successively, and serving multiple persons simultaneously. Though the relationship between device and people is characterized in all variants and the approach has several identifier, the life cycle itself is not regarded [REQ11]. It is furthermore unknown how the approach should interact with other approaches and protocols [REQ3].

\subsection{User-Centric Identity Management with UMA}

The works published by Kumar et al.~\cite{8442455} and Slomovic~\cite{6837386} make it clear that pseudonymity and anonymity can be important in certain use cases. Identities from different social networks can be aggregated and lead to one person. UMA manages to decouple identity resolution from the maintenance of identity information~\cite{6822222}
\cite{10.1145/1866855.1866865} \cite{7163222} and even one application for IoT by Cruz-Piris et al.~\cite{umaiot} is published. Although UMA is based on OAuth, the principle can be applied to SAML as well~\cite{samluma}. Nevertheless, the principle of UMA might not be possible to apply to all use cases [REQ3].

\subsection{User-Centric Identity Management with SSI}

The principles of SSI management are described by Toth and Anderson-Priddy~\cite{8713271}. Ferdous et al.~\cite{8776589} broaden the laws of identities from existence, autonomy, ownership, access, single source, protection, availability, and persistence to user control and consent, minimal disclosure for a constrained use, justifiable parties, directed identity, pluralism of operators and technologies, human integration, and consistent experience across contexts for self-sovereign identities. This step is seen as an evolution of identity management by Sovrin~\cite{sovrin}.

SSIs are typically implemented by blockchain, which have decentralized identifiers (DIDs), proposed by the W3C, as a new type of identifier for verifiable, decentralized digital identities. Claim Chain by Stokkink and Pouwelse~\cite{8726562} is a pure self-sovereign building on claim model, which is practically used, but also rather restrictive. No data is sent without consent, but also at the same time only attested attributes are possible [REQ3]. DecentID by Friebe et al.~\cite{8455884} and Steichen et al.~\cite{8726493} is based on Ethereum smart contracts and tries to preserve the privacy of the attributes. The approach utilizes external storage like distributed hash tables (DHTs) for attributes for scalability reasons. The nodes do not mine efficiently with large files [REQ4]. Additionally, a linkage between identities to centralized registration contracts is needed [REQ9]. Sora Identity~\cite{8377927} uses the mobile phone of the user as one of the main actors. Though this approach preserves the privacy, the life cycle of a device and the usage of different devices can get problematic [REQ8]. Another blockchain-based approach combines it with PKI~\cite{8406325} \cite{secrypt17} [REQ3]. The approaches support the revocation of certificates, but have scalability problems [REQ4]. The Horcux Protocol~\cite{8489316}\cite{Othman2019} is a decentralized biometric credential storage using SSI with Biometric Open Protocol Standard (BOPS). This is done although biometrics tend to have false-positives and false-negatives and their implementations are often broken [REQ7]. In~\cite{schanzenbach2016}, Schanzenbach and Banse describe an approach, which is adapted from UMA applying NameID~\cite{nameid}. As they use a secure name system - GNU Name System (GNS) - to store data, they do not require a global, public ledger and do not compromise the privacy of users.  Ahadipour and Schanzenbach~\cite{ahadipo2017} present a survey on authorization in distributed systems.  In~\cite{schanzenbach2018abd}, the authors describe the decentralized attribute-based delegation (ABD). reclaimID by Schanzenbach et al.~\cite{8456003} is based on the work published beforehand. Attributes are encrypted by attribute-based encryption (ABE), which means the attributes are self-signed with the private key associated with the identity of the user. The approach allows the use of one or more identities with different attributes. Nevertheless, privacy might be one drawback as well as trust into the attributes [REQ6] [REQ9].

\subsection{User-Centric Identity Management with SSI for Cloud Environments}

The identity wallet platform CREDENTIAL~\cite{Veseli:2019:EPD:3297280.3297429} \cite{Kostopoulos:2017:TAS:3098954.3104061} \cite{Karegar2016} \cite{7784641} aims at the development of a secure and privacy-preserving data sharing platform that puts the user under full control. \cite{7784641} shows how proxy-re-encryption and redactable signature technology can be integrated into their workflows, which is basically a man in the middle. The platform has three main actors: wallet platform, user, and the data receiver, which can be either user or SP, as described in~\cite{Karegar2016}. A metadata preserving high-level architecture is proposed, which might be practically challenging due to log files leaking information [REQ7]. Kostopoulos et al.~\cite{Kostopoulos:2017:TAS:3098954.3104061} describe the architecture of CREDENTIAL. Interfaces to existing protocols are missing [REQ3] [REQ11]. Finally, Veseli et al.~\cite{Veseli:2019:EPD:3297280.3297429} focus on privacy by design. The LINDDUN (Linkability, Identifiability, Non-repudiation, Detectability, information Disclosure, content Unawareness, and policy and consent Non-compliance) method~\cite{Deng2011}, a privacy threat analysis framework, was used to establish privacy by design. Thereby, the data is encrypted, though the approach is cloud-based. The user has to trust the wallet provider [REQ3] [REQ6] [REQ11].

\subsection{User-Centric Identity Management with SSI as Federation}

NEXTLEAP by Halpin et al.~\cite{Halpin:2017:NDI:3098954.3104056} is an approach for FIM with blind signatures based on MACs to improve OIDC. The identity itself is a privacy-enhanced based blockchain identity with a decentralized PKI. Therefore, access control is performed in a decentralized manner. Though the purpose is secure messaging, the scalability can be a drawback in large-scale infrastructures [REQ4]. \cite{8422733} is designed with having a cross-organizational authentication system in mind. One-way hashed enforce the concept of one-time pad passwords, which can provide unlinkability. Unfortunately, the performance is not clearly described [REQ4].

\subsection{User-Centric Identity Management with SSI as Cloud Federation}

The goal of the EU project SUNFISH was the establishment of a cloud federation, q.\,v. Alansari et al.~\cite{8030666} \cite{7980160}. The cloud federation is a collaboration of organizations sharing data hosted on their private cloud infrastructures for business opportunity. While the approach introduces distributed access in a privacy-preserving way, the user must be identified and authenticated beforehand. eIDAS can be used as a source of trust, though the trust within a cloud federation is not clearly described [REQ6]. Although the anonymity for users by the usage of tokens instead of attributes sounds promising, tokens might not in every setting. Additionally, the approach is solely based on Intel Software Guard Extensions (SGX) [REQ11]. The approach of dynamic cloud federation by Bendiab evolves from a dynamic cloud federation to blockchain with fuzzy cognitive maps dynamic trust model~\cite{Keltoum:2017:DFI:3018896.3025152} \cite{cloudtrust} \cite{trustcloud} \cite{dynamictrust}. In~\cite{Keltoum:2017:DFI:3018896.3025152}, the authors describe the basic approach with the entities user, cloud service provider (CSP), IdP, federation provider (FP), and certificate authority. The token is sent from one FP to another FP for validation and transformation. The CSP verifies the token's legitimacy, before the user can access the cloud service. This could also be done by a simple proxy between two protocols, like SAML and OIDC in eduGAIN [REQ4]. Trust is described in~\cite{cloudtrust}. Each user gets his identity from a trusted IdP. If the CSP does not trust this IdP, the trust can be computed for each other in real time. Based on the result, a connection is established. The authors describe a complex trust evaluation, which is derived from different characteristics, like security, and privacy, though these factors are difficult to measure and access [REQ6]. The trust is then put into blockchain~\cite{trustcloud}. The CSP can manage the trust relationships in a dynamic and distributed manner without the need for centralized authorities, like IdP. Nevertheless, a Trust Management Platform (TMP) is needed and the CSPs are responsible for authenticating their registered users. \cite{dynamictrust} is the next step with fuzzy cognitive maps for modeling and evaluating trust relationships. It uses a trust computation model, similar to the characteristics introduced beforehand. It can calculate the trust of unknown entities, if further information are available. Unfortunately, LoA methods like Vectors of Trust (VoT)~\cite{RFC8485} are not taken into account [REQ6] [REQ11]. The approach eases the creation of secure Infrastructure as a Service (IaaS) cloud federation. Though the approach makes use of anonymity approaches, it still stores personal information in an external database [REQ9]. Additionally, the approach concentrates on one use case [REQ3].

\subsection{Conclusion from User-Centric Identity Management}

A summary of user-centric identity management approaches is shown in Table~\ref{tab:ucim}. UCIM has drawbacks in Interoperability [REQ3], Functionality [REQ5], and Liability [REQ10], though there are differences between older approaches, UMA, and SSI. Within SSI, it partly depends on use cases, centralized / decentralized management, among others. The presented approaches for UMA miss Interoperability [REQ3]. The principle of UMA should work for many use cases, although this is not further explored to the knowledge of the authors. UCIM for IoT has issues with Integration [REQ11] and presumably with Interoperability [REQ3]. Looking at the presented SSI approaches, Scalability [REQ4] and Privacy [REQ9] are often stated. Many approaches are tested within smaller environments, leading to problems with scalability. Privacy seems contradictory as the approaches focus on users. Not all approaches concentrate on privacy and others seem to have issues as well. Focusing on cloud and federation use cases, Scalability [REQ4], Trustworthiness [REQ6], and Integration [REQ11] are often missing. Trustworthiness is difficult, as the user has more control over his data while the service provider needs to trust the user's attributes.

\begin{table}
\caption{Overview of Approaches for User-Centric Identity Management}
\label{tab:ucim}
\begin{tabular}{p{1.2cm}p{2.5cm}p{3.5cm}}
Ref & Name & Requirements not fulfilled  \\ \hline
 & UCIM & [REQ3] [REQ5] [REQ10] \\
\cite{7021724} & UCIM for IoT & [REQ3] [REQ11] \\
\cite{6822222} & UMA & [REQ3] \\
\cite{umaiot} & UMA for IoT & [REQ3] \\
\cite{8726562} & Claim Chain & [REQ3] \\
\cite{8455884} & DecentID & [REQ4] [REQ9] \\
\cite{8726493} & SSI for IPFS & [REQ4] [REQ9] \\
\cite{8377927} & Sora Identity & [REQ8] \\
\cite{8406325} \cite{secrypt17}  & Blockchain-based PKI management framework & [REQ3] [REQ4] \\
\cite{8489316} \cite{Othman2019} & Horcux Protocol & [REQ7] \\
\cite{8456003} & reclaimID & [REQ6] [REQ9] \\
\cite{Veseli:2019:EPD:3297280.3297429} & CREDENTIAL &  [REQ3] [REQ6] [REQ7] [REQ11] \\
\cite{Halpin:2017:NDI:3098954.3104056} & NEXTLEAP & [REQ4] \\
\cite{8422733}  & Blockchain-based AAA system & [REQ4] \\
\cite{8030666} \cite{7980160} & SUNFISH & [REQ6] [REQ11] \\
\cite{Keltoum:2017:DFI:3018896.3025152} & Dynamic cloud federation & [REQ3] [REQ4] [REQ6] [REQ9] [REQ11] \\
\end{tabular}
\end{table}

\section{Conclusion and Future Work}\label{sec:summary}

Nowadays, security is one key requirement in a network environment. Identities are everywhere and with the growing identity theft, secure identity management gets more relevant than before. In this paper, we extend known requirements. Based on new approaches as well as the heterogeneous environments and possible use cases, new requirements were added. At the same time, others were newly defined. Management [REQ1] of identities is a new requirement, while Functionality [REQ5] is still needed for interoperability between all these approaches. The GDPR leads to the change from Mobility to Portability [REQ8], as users can request to port their account and more and more different devices are used. This also leads into the division Scalability [REQ4] and Integration [REQ11], which was originally Affordability.

As a next step, we give a broad overview of different identity management approaches structured in identity management models and use cases. This ranges  from centralized identity management for specific use cases, over federated identity management, to user-centric approaches, like user-managed access and self-sovereign identities.  We compare these approaches with the derived requirements, which gives us an overview of advantages as well as drawbacks.

Centralized approaches lack Interoperability [REQ3], Trustworthiness [REQ6], Management [REQ1], and Integration [REQ11]. Portability [REQ8] cannot be estimated. The shortcomings are inherent of the model. Federated Identity Management has general issues with Interoperability [REQ3], Functionality [REQ5], and Liability [REQ10]. SAML misses Scalability [REQ4] and Portability [REQ8] as well. Trustworthiness [REQ6] and Integration [REQ11] are additional shortcomings of the described FIM approaches. User-Centric Identity Management has drawbacks in Interoperability [REQ3], Functionality [REQ5], and Liability [REQ10]. Also Trustworthiness [REQ6] is an issue. Depending on the approach, Scalability [REQ4] and Integration [REQ11] are unavailable as well. No approach has managed to meet all requirements. Two requirements were out of scope of the described approaches: Liability [REQ10] and Functionality [REQ5]. Liability is mainly endorsed by the home organization, while additional approaches try to fulfill Functionality [REQ5]. Interoperability [REQ3] and Trustworthiness [REQ6] might be met within the approach or model but not crosswise. Thereby, Integration [REQ11] is also difficult. Simply combining different approaches does not satisfy the stated requirements. Running several approaches in parallel binds more manpower and time. Additionally, it can lead to different policies and further inconsistent database. Therefore, at least an overview and connectors between the approaches are needed to interact between the approaches and provide an overview. In order to meet all requirements, further work is needed.

As a result, stakeholders would profit from a framework including different models and, thereby, approaches. The framework needs to be flexible enough for future work. In order to meet the missing requirements, it needs to comprise of additional tools and recommended processes, including connectors between different approaches and an overview. For future work, we plan to develop missing components in the design of a holistic framework. The framework will be adapted for different use cases in a further step. Additionally, a categorization for identity management models will be created.

\bibliographystyle{ACM-Reference-Format}
\bibliography{sota-arxiv}


\begin{thebibliography}{88}


\ifx \showCODEN    \undefined \def \showCODEN     #1{\unskip}     \fi
\ifx \showDOI      \undefined \def \showDOI       #1{#1}\fi
\ifx \showISBNx    \undefined \def \showISBNx     #1{\unskip}     \fi
\ifx \showISBNxiii \undefined \def \showISBNxiii  #1{\unskip}     \fi
\ifx \showISSN     \undefined \def \showISSN      #1{\unskip}     \fi
\ifx \showLCCN     \undefined \def \showLCCN      #1{\unskip}     \fi
\ifx \shownote     \undefined \def \shownote      #1{#1}          \fi
\ifx \showarticletitle \undefined \def \showarticletitle #1{#1}   \fi
\ifx \showURL      \undefined \def \showURL       {\relax}        \fi
\providecommand\bibfield[2]{#2}
\providecommand\bibinfo[2]{#2}
\providecommand\natexlab[1]{#1}
\providecommand\showeprint[2][]{arXiv:#2}

\bibitem[\protect\citeauthoryear{Ahadipour and Schanzenbach}{Ahadipour and
  Schanzenbach}{2017}]%
        {ahadipo2017}
\bibfield{author}{\bibinfo{person}{Ava Ahadipour} {and} \bibinfo{person}{Martin
  Schanzenbach}.} \bibinfo{year}{2017}\natexlab{}.
\newblock \showarticletitle{{ A Survey on Authorization in Distributed Systems:
  Information Storage, Data Retrieval and Trust Evaluation}}. In
  \bibinfo{booktitle}{\emph{16th IEEE International Conference on Trust,
  Security and Privacy in Computing and Communications,}}. IEEE,
  \bibinfo{pages}{1016--1023}.
\newblock
\urldef\tempurl%
\url{http://dx.doi.org/10.1109/Trustcom/BigDataSE/ICESS.2017.346}
\showURL{%
\tempurl}


\bibitem[\protect\citeauthoryear{{Alansari}, {Paci}, {Margheri}, and
  {Sassone}}{{Alansari} et~al\mbox{.}}{2017b}]%
        {8030666}
\bibfield{author}{\bibinfo{person}{Shorouq {Alansari}},
  \bibinfo{person}{Federica {Paci}}, \bibinfo{person}{Andrea {Margheri}}, {and}
  \bibinfo{person}{Vladimiro {Sassone}}.} \bibinfo{year}{2017}\natexlab{b}.
\newblock \showarticletitle{{Privacy-Preserving Access Control in Cloud
  Federations}}. In \bibinfo{booktitle}{\emph{2017 IEEE 10th International
  Conference on Cloud Computing (CLOUD)}}. \bibinfo{pages}{757--760}.
\newblock
\showISSN{2159-6190}
\urldef\tempurl%
\url{https://doi.org/10.1109/CLOUD.2017.108}
\showDOI{\tempurl}


\bibitem[\protect\citeauthoryear{{Alansari}, {Paci}, and {Sassone}}{{Alansari}
  et~al\mbox{.}}{2017a}]%
        {7980160}
\bibfield{author}{\bibinfo{person}{Shorouq {Alansari}},
  \bibinfo{person}{Federica {Paci}}, {and} \bibinfo{person}{Vladimiro
  {Sassone}}.} \bibinfo{year}{2017}\natexlab{a}.
\newblock \showarticletitle{{A Distributed Access Control System for Cloud
  Federations}}. In \bibinfo{booktitle}{\emph{2017 IEEE 37th International
  Conference on Distributed Computing Systems (ICDCS)}}.
  \bibinfo{pages}{2131--2136}.
\newblock
\showISSN{1063-6927}
\urldef\tempurl%
\url{https://doi.org/10.1109/ICDCS.2017.241}
\showDOI{\tempurl}


\bibitem[\protect\citeauthoryear{Alonso, Pozo, Choque, Bueno, {Salvach{\'u}a},
  Diez, {Mar{\`i}n}, and Alonso}{Alonso et~al\mbox{.}}{2019}]%
        {8754671}
\bibfield{author}{\bibinfo{person}{Alvaro Alonso}, \bibinfo{person}{Alejandro
  Pozo}, \bibinfo{person}{Johnny Choque}, \bibinfo{person}{Gloria Bueno},
  \bibinfo{person}{Joaquin {Salvach{\'u}a}}, \bibinfo{person}{Luis Diez},
  \bibinfo{person}{Jorge {Mar{\`i}n}}, {and} \bibinfo{person}{Pedro Luis~Chas
  Alonso}.} \bibinfo{year}{2019}\natexlab{}.
\newblock \showarticletitle{{An Identity Framework for Providing Access to
  FIWARE OAuth 2.0 - Based Services According to the eIDAS European
  Regulation}}.
\newblock \bibinfo{journal}{\emph{{IEEE Access}}} (\bibinfo{year}{2019}),
  \bibinfo{pages}{88435--88449}.
\newblock
\showISSN{2169-3536}
\urldef\tempurl%
\url{https://doi.org/10.1109/ACCESS.2019.2926556}
\showDOI{\tempurl}


\bibitem[\protect\citeauthoryear{{Alrodhan} and {Mitchell}}{{Alrodhan} and
  {Mitchell}}{2007}]%
        {4299788}
\bibfield{author}{\bibinfo{person}{W.~A. {Alrodhan}} {and}
  \bibinfo{person}{C.~J. {Mitchell}}.} \bibinfo{year}{2007}\natexlab{}.
\newblock \showarticletitle{Addressing privacy issues in CardSpace}. In
  \bibinfo{booktitle}{\emph{Third International Symposium on Information
  Assurance and Security}}. \bibinfo{pages}{285--291}.
\newblock
\showISSN{null}
\urldef\tempurl%
\url{https://doi.org/10.1109/IAS.2007.12}
\showDOI{\tempurl}


\bibitem[\protect\citeauthoryear{Arias~Cabarcos, Almen{\'a}rez,
  G{\'o}mez~M{\'a}rmol, and Mar{\'i}n}{Arias~Cabarcos et~al\mbox{.}}{2013}]%
        {Arias:2013:IdMRep}
\bibfield{author}{\bibinfo{person}{Patricia Arias~Cabarcos},
  \bibinfo{person}{Florina Almen{\'a}rez}, \bibinfo{person}{F{\'e}lix
  G{\'o}mez~M{\'a}rmol}, {and} \bibinfo{person}{Andr{\'e}s Mar{\'i}n}.}
  \bibinfo{year}{2013}\natexlab{}.
\newblock \showarticletitle{{To Federate or Not To Federate: A Reputation-Based
  Mechanism to Dynamize Cooperation in Identity Management}}.
\newblock \bibinfo{journal}{\emph{Wireless Personal Communications}}
  (\bibinfo{year}{2013}), \bibinfo{pages}{1--18}.
\newblock
\showISSN{0929-6212}
\urldef\tempurl%
\url{https://doi.org/10.1007/s11277-013-1338-y}
\showDOI{\tempurl}


\bibitem[\protect\citeauthoryear{Axon and Goldsmith}{Axon and
  Goldsmith}{2017}]%
        {secrypt17}
\bibfield{author}{\bibinfo{person}{Louise Axon} {and} \bibinfo{person}{Michael
  Goldsmith}.} \bibinfo{year}{2017}\natexlab{}.
\newblock \showarticletitle{{PB-PKI: A Privacy-aware Blockchain-based PKI}}. In
  \bibinfo{booktitle}{\emph{Proceedings of the 14th International Joint
  Conference on e-Business and Telecommunications - Volume 6: SECRYPT, (ICETE
  2017)}}. INSTICC, \bibinfo{publisher}{SciTePress}, \bibinfo{pages}{311--318}.
\newblock
\showISBNx{978-989-758-259-2}
\urldef\tempurl%
\url{https://doi.org/10.5220/0006419203110318}
\showDOI{\tempurl}


\bibitem[\protect\citeauthoryear{Bargh, Hulsebosch, and Zandbelt}{Bargh
  et~al\mbox{.}}{2011}]%
        {MortazaS.Bargh2011}
\bibfield{author}{\bibinfo{person}{Mortaza~S. Bargh}, \bibinfo{person}{Bob
  Hulsebosch}, {and} \bibinfo{person}{Hans Zandbelt}.}
  \bibinfo{year}{2011}\natexlab{}.
\newblock \showarticletitle{Scalability of trust and metadata exchange across
  federations}.
\newblock \bibinfo{journal}{\emph{TNC}} (\bibinfo{year}{2011}).
\newblock


\bibitem[\protect\citeauthoryear{{Beltran} and {Skarmeta}}{{Beltran} and
  {Skarmeta}}{2016}]%
        {7845482}
\bibfield{author}{\bibinfo{person}{V. {Beltran}} {and} \bibinfo{person}{A.~F.
  {Skarmeta}}.} \bibinfo{year}{2016}\natexlab{}.
\newblock \showarticletitle{{An overview on delegated authorization for CoAP:
  Authentication and authorization for Constrained Environments (ACE)}}. In
  \bibinfo{booktitle}{\emph{{2016 IEEE 3rd World Forum on Internet of Things
  (WF-IoT)}}}. \bibinfo{pages}{706--710}.
\newblock
\showISSN{null}
\urldef\tempurl%
\url{https://doi.org/10.1109/WF-IoT.2016.7845482}
\showDOI{\tempurl}


\bibitem[\protect\citeauthoryear{Bendiab, Kolokotronis, Shiaeles, and
  Boucherkha}{Bendiab et~al\mbox{.}}{2018a}]%
        {trustcloud}
\bibfield{author}{\bibinfo{person}{Gueltoum Bendiab}, \bibinfo{person}{Nicholas
  Kolokotronis}, \bibinfo{person}{Stavros Shiaeles}, {and}
  \bibinfo{person}{Samia Boucherkha}.} \bibinfo{year}{2018}\natexlab{a}.
\newblock \showarticletitle{{WiP: A Novel Blockchain-Based Trust Model for
  Cloud Identity Management}}. In \bibinfo{booktitle}{\emph{2018 IEEE 16th Intl
  Conf on Dependable, Autonomic and Secure Computing, 16th Intl Conf on
  Pervasive Intelligence and Computing, 4th Intl Conf on Big Data Intelligence
  and Computing and Cyber Science and Technology
  Congress(DASC/PiCom/DataCom/CyberSciTech)}}. \bibinfo{pages}{724--729}.
\newblock
\urldef\tempurl%
\url{https://doi.org/10.1109/DASC/PiCom/DataCom/CyberSciTec.2018.00126}
\showDOI{\tempurl}


\bibitem[\protect\citeauthoryear{Bendiab, Shiaeles, and Boucherkha}{Bendiab
  et~al\mbox{.}}{2018b}]%
        {cloudtrust}
\bibfield{author}{\bibinfo{person}{Gueltoum Bendiab}, \bibinfo{person}{Stavros
  Shiaeles}, {and} \bibinfo{person}{Samia Boucherkha}.}
  \bibinfo{year}{2018}\natexlab{b}.
\newblock \showarticletitle{{A New Dynamic Trust Model for "On Cloud" Federated
  Identity Management}}. In \bibinfo{booktitle}{\emph{2018 9th IFIP
  International Conference on New Technologies, Mobility and Security (NTMS)}}.
  \bibinfo{pages}{1--5}.
\newblock
\urldef\tempurl%
\url{https://doi.org/10.1109/NTMS.2018.8328673}
\showDOI{\tempurl}


\bibitem[\protect\citeauthoryear{Bendiab, Shiaeles, Boucherkha, and
  Ghita}{Bendiab et~al\mbox{.}}{2019}]%
        {dynamictrust}
\bibfield{author}{\bibinfo{person}{Gueltoum Bendiab}, \bibinfo{person}{Stavros
  Shiaeles}, \bibinfo{person}{Samia Boucherkha}, {and} \bibinfo{person}{B.V.
  Ghita}.} \bibinfo{year}{2019}\natexlab{}.
\newblock \showarticletitle{{FCMDT: A Novel Fuzzy Cognitive Maps Dynamic Trust
  Model for Cloud Federated Identity Management}}.
\newblock \bibinfo{journal}{\emph{Computers \& Security}} (\bibinfo{date}{06}
  \bibinfo{year}{2019}).
\newblock
\urldef\tempurl%
\url{https://doi.org/10.1016/j.cose.2019.06.011}
\showDOI{\tempurl}


\bibitem[\protect\citeauthoryear{{Berbecaru}, {Atzeni}, {Benedictis}, and
  {Smiraglia}}{{Berbecaru} et~al\mbox{.}}{2017}]%
        {7912677}
\bibfield{author}{\bibinfo{person}{Diana {Berbecaru}}, \bibinfo{person}{Andrea
  {Atzeni}}, \bibinfo{person}{Marco~de {Benedictis}}, {and}
  \bibinfo{person}{Paolo {Smiraglia}}.} \bibinfo{year}{2017}\natexlab{}.
\newblock \showarticletitle{{Towards Stronger Data Security in an eID
  Management Infrastructure}}. In \bibinfo{booktitle}{\emph{2017 25th Euromicro
  International Conference on Parallel, Distributed and Network-based
  Processing (PDP)}}. \bibinfo{pages}{391--395}.
\newblock
\showISSN{2377-5750}
\urldef\tempurl%
\url{https://doi.org/10.1109/PDP.2017.90}
\showDOI{\tempurl}


\bibitem[\protect\citeauthoryear{{Berbecaru} and {Lioy}}{{Berbecaru} and
  {Lioy}}{2018}]%
        {8540674}
\bibfield{author}{\bibinfo{person}{Diana {Berbecaru}} {and}
  \bibinfo{person}{Antonio {Lioy}}.} \bibinfo{year}{2018}\natexlab{}.
\newblock \showarticletitle{{On integration of academic attributes in the eIDAS
  infrastructure to support cross-border services}}. In
  \bibinfo{booktitle}{\emph{2018 22nd International Conference on System
  Theory, Control and Computing (ICSTCC)}}. \bibinfo{pages}{691--696}.
\newblock
\showISSN{2372-1618}
\urldef\tempurl%
\url{https://doi.org/10.1109/ICSTCC.2018.8540674}
\showDOI{\tempurl}


\bibitem[\protect\citeauthoryear{Berbecaru, Lioy, and Cameroni}{Berbecaru
  et~al\mbox{.}}{2019}]%
        {info10060210}
\bibfield{author}{\bibinfo{person}{Diana Berbecaru}, \bibinfo{person}{Antonio
  Lioy}, {and} \bibinfo{person}{Cesare Cameroni}.}
  \bibinfo{year}{2019}\natexlab{}.
\newblock \showarticletitle{{Electronic Identification for Universities:
  Building Cross-Border Services Based on the eIDAS Infrastructure}}.
\newblock \bibinfo{journal}{\emph{Information}} \bibinfo{volume}{10},
  \bibinfo{number}{6} (\bibinfo{year}{2019}).
\newblock
\showISSN{2078-2489}
\urldef\tempurl%
\url{https://doi.org/10.3390/info10060210}
\showDOI{\tempurl}


\bibitem[\protect\citeauthoryear{{Boujezza}, {AL-Mufti}, {Ayed}, and
  {Saidane}}{{Boujezza} et~al\mbox{.}}{2015}]%
        {7507266}
\bibfield{author}{\bibinfo{person}{H. {Boujezza}}, \bibinfo{person}{M.
  {AL-Mufti}}, \bibinfo{person}{H.~K.~B. {Ayed}}, {and} \bibinfo{person}{L.
  {Saidane}}.} \bibinfo{year}{2015}\natexlab{}.
\newblock \showarticletitle{A taxonomy of identities management systems in
  IOT}. In \bibinfo{booktitle}{\emph{2015 IEEE/ACS 12th International
  Conference of Computer Systems and Applications (AICCSA)}}.
  \bibinfo{pages}{1--8}.
\newblock
\showISSN{2161-5330}
\urldef\tempurl%
\url{https://doi.org/10.1109/AICCSA.2015.7507266}
\showDOI{\tempurl}


\bibitem[\protect\citeauthoryear{Cruz-Piris, Rivera, Marsa-Maestre, De~la Hoz,
  and Velasco}{Cruz-Piris et~al\mbox{.}}{2018}]%
        {umaiot}
\bibfield{author}{\bibinfo{person}{Luis Cruz-Piris}, \bibinfo{person}{Diego
  Rivera}, \bibinfo{person}{Ivan Marsa-Maestre}, \bibinfo{person}{Enrique De~la
  Hoz}, {and} \bibinfo{person}{Juan~R. Velasco}.}
  \bibinfo{year}{2018}\natexlab{}.
\newblock \showarticletitle{{Access Control Mechanism for IoT Environments
  Based on Modelling Communication Procedures as Resources}}.
\newblock \bibinfo{journal}{\emph{Sensors}} \bibinfo{volume}{18},
  \bibinfo{number}{3} (\bibinfo{date}{Mar} \bibinfo{year}{2018}),
  \bibinfo{pages}{917}.
\newblock
\showISSN{1424-8220}
\urldef\tempurl%
\url{https://doi.org/10.3390/s18030917}
\showDOI{\tempurl}


\bibitem[\protect\citeauthoryear{Deng, Wuyts, Scandariato, Preneel, and
  Joosen}{Deng et~al\mbox{.}}{2011}]%
        {Deng2011}
\bibfield{author}{\bibinfo{person}{Mina Deng}, \bibinfo{person}{Kim Wuyts},
  \bibinfo{person}{Riccardo Scandariato}, \bibinfo{person}{Bart Preneel}, {and}
  \bibinfo{person}{Wouter Joosen}.} \bibinfo{year}{2011}\natexlab{}.
\newblock \showarticletitle{{A privacy threat analysis framework: supporting
  the elicitation and fulfillment of privacy requirements}}.
\newblock \bibinfo{journal}{\emph{Requirements Engineering}}
  \bibinfo{volume}{16}, \bibinfo{number}{1} (\bibinfo{date}{Mar}
  \bibinfo{year}{2011}), \bibinfo{pages}{3--32}.
\newblock
\showISSN{1432-010X}
\urldef\tempurl%
\url{https://doi.org/10.1007/s00766-010-0115-7}
\showDOI{\tempurl}


\bibitem[\protect\citeauthoryear{{Dhungana}, {Mohammad}, {Sharma}, and
  {Schoen}}{{Dhungana} et~al\mbox{.}}{2013}]%
        {6544386}
\bibfield{author}{\bibinfo{person}{R.~D. {Dhungana}}, \bibinfo{person}{A.
  {Mohammad}}, \bibinfo{person}{A. {Sharma}}, {and} \bibinfo{person}{I.
  {Schoen}}.} \bibinfo{year}{2013}\natexlab{}.
\newblock \showarticletitle{{Identity Management Framework for Cloud Networking
  Infrastructure}}. In \bibinfo{booktitle}{\emph{{2013 9th International
  Conference on Innovations in Information Technology (IIT)}}}.
  \bibinfo{pages}{13--17}.
\newblock
\showISSN{null}
\urldef\tempurl%
\url{https://doi.org/10.1109/Innovations.2013.6544386}
\showDOI{\tempurl}


\bibitem[\protect\citeauthoryear{Ferdous and Poet}{Ferdous and Poet}{2013}]%
        {dynfed}
\bibfield{author}{\bibinfo{person}{Md.Sadek Ferdous} {and} \bibinfo{person}{Ron
  Poet}.} \bibinfo{year}{2013}\natexlab{}.
\newblock \showarticletitle{{Dynamic Identity Federation Using Security
  Assertion Markup Language (SAML)}}.
\newblock In \bibinfo{booktitle}{\emph{Policies and Research in Identity
  Management}}, \bibfield{editor}{\bibinfo{person}{Simone Fischer-H{\"u}bner},
  \bibinfo{person}{Elisabeth de~Leeuw}, {and} \bibinfo{person}{Chris Mitchell}}
  (Eds.). \bibinfo{series}{IFIP Advances in Information and Communication
  Technology}, Vol.~\bibinfo{volume}{396}. \bibinfo{publisher}{Springer Berlin
  Heidelberg}, \bibinfo{pages}{131--146}.
\newblock
\urldef\tempurl%
\url{https://doi.org/10.1007/978-3-642-37282-7_13}
\showDOI{\tempurl}


\bibitem[\protect\citeauthoryear{Ferdous, Chowdhury, and Alassafi}{Ferdous
  et~al\mbox{.}}{2019}]%
        {8776589}
\bibfield{author}{\bibinfo{person}{Md~Sadek Ferdous}, \bibinfo{person}{Farida
  Chowdhury}, {and} \bibinfo{person}{Madini~O. Alassafi}.}
  \bibinfo{year}{2019}\natexlab{}.
\newblock \showarticletitle{{In Search of Self-Sovereign Identity Leveraging
  Blockchain Technology}}.
\newblock \bibinfo{journal}{\emph{IEEE Access}}  \bibinfo{volume}{7}
  (\bibinfo{year}{2019}), \bibinfo{pages}{103059--103079}.
\newblock
\showISSN{2169-3536}
\urldef\tempurl%
\url{https://doi.org/10.1109/ACCESS.2019.2931173}
\showDOI{\tempurl}


\bibitem[\protect\citeauthoryear{{Ferdous} and {Poet}}{{Ferdous} and
  {Poet}}{2012}]%
        {6266958}
\bibfield{author}{\bibinfo{person}{M.~S. {Ferdous}} {and} \bibinfo{person}{R.
  {Poet}}.} \bibinfo{year}{2012}\natexlab{}.
\newblock \showarticletitle{A comparative analysis of Identity Management
  Systems}. In \bibinfo{booktitle}{\emph{2012 International Conference on High
  Performance Computing Simulation (HPCS)}}. \bibinfo{pages}{454--461}.
\newblock
\showISSN{null}
\urldef\tempurl%
\url{https://doi.org/10.1109/HPCSim.2012.6266958}
\showDOI{\tempurl}


\bibitem[\protect\citeauthoryear{{Florea}, {Rughinis}, {Ruse}, and
  {Dragomir}}{{Florea} et~al\mbox{.}}{2017}]%
        {7968561}
\bibfield{author}{\bibinfo{person}{I. {Florea}}, \bibinfo{person}{R.
  {Rughinis}}, \bibinfo{person}{L. {Ruse}}, {and} \bibinfo{person}{D.
  {Dragomir}}.} \bibinfo{year}{2017}\natexlab{}.
\newblock \showarticletitle{{Survey of Standardized Protocols for the Internet
  of Things}}. In \bibinfo{booktitle}{\emph{{2017 21st International Conference
  on Control Systems and Computer Science (CSCS)}}}. \bibinfo{pages}{190--196}.
\newblock
\showISSN{2379-0482}
\urldef\tempurl%
\url{https://doi.org/10.1109/CSCS.2017.33}
\showDOI{\tempurl}


\bibitem[\protect\citeauthoryear{{Friebe}, {Sobik}, and {Zitterbart}}{{Friebe}
  et~al\mbox{.}}{2018}]%
        {8455884}
\bibfield{author}{\bibinfo{person}{Sebastian {Friebe}}, \bibinfo{person}{Ingo
  {Sobik}}, {and} \bibinfo{person}{Martina {Zitterbart}}.}
  \bibinfo{year}{2018}\natexlab{}.
\newblock \showarticletitle{{DecentID: Decentralized and Privacy-Preserving
  Identity Storage System Using Smart Contracts}}. In
  \bibinfo{booktitle}{\emph{2018 17th IEEE International Conference On Trust,
  Security And Privacy In Computing And Communications/ 12th IEEE International
  Conference On Big Data Science And Engineering (TrustCom/BigDataSE)}}.
  \bibinfo{pages}{37--42}.
\newblock
\showISSN{2324--9013}
\urldef\tempurl%
\url{https://doi.org/10.1109/TrustCom/BigDataSE.2018.00016}
\showDOI{\tempurl}


\bibitem[\protect\citeauthoryear{{Gao}, {Zhang}, {Xia}, and {Ma}}{{Gao}
  et~al\mbox{.}}{2016}]%
        {7423325}
\bibfield{author}{\bibinfo{person}{F. {Gao}}, \bibinfo{person}{F. {Zhang}},
  \bibinfo{person}{J. {Xia}}, {and} \bibinfo{person}{Z. {Ma}}.}
  \bibinfo{year}{2016}\natexlab{}.
\newblock \showarticletitle{General identity management model for big data
  analysis}. In \bibinfo{booktitle}{\emph{2016 18th International Conference on
  Advanced Communication Technology (ICACT)}}. \bibinfo{pages}{1--1}.
\newblock
\showISSN{null}
\urldef\tempurl%
\url{https://doi.org/10.1109/ICACT.2016.7423325}
\showDOI{\tempurl}


\bibitem[\protect\citeauthoryear{G{\'E}ANT}{G{\'E}ANT}{2016}]%
        {edugainstatus}
\bibfield{author}{\bibinfo{person}{G{\'E}ANT}.}
  \bibinfo{year}{2016}\natexlab{}.
\newblock \bibinfo{title}{eduGAIN technical site}.
\newblock
  \bibinfo{howpublished}{\url{https://technical.edugain.org/status.php}}.
\newblock
\newblock
\shownote{[Online, \today].}


\bibitem[\protect\citeauthoryear{{Haddouti} and {Ech-Cherif El
  Kettani}}{{Haddouti} and {Ech-Cherif El Kettani}}{2019}]%
        {8742375}
\bibfield{author}{\bibinfo{person}{Samia~El {Haddouti}} {and}
  \bibinfo{person}{M.~Dafir {Ech-Cherif El Kettani}}.}
  \bibinfo{year}{2019}\natexlab{}.
\newblock \showarticletitle{{Analysis of Identity Management Systems Using
  Blockchain Technology}}. In \bibinfo{booktitle}{\emph{2019 International
  Conference on Advanced Communication Technologies and Networking (CommNet)}}.
  \bibinfo{pages}{1--7}.
\newblock
\urldef\tempurl%
\url{https://doi.org/10.1109/COMMNET.2019.8742375}
\showDOI{\tempurl}


\bibitem[\protect\citeauthoryear{Halpin}{Halpin}{2017}]%
        {Halpin:2017:NDI:3098954.3104056}
\bibfield{author}{\bibinfo{person}{Harry Halpin}.}
  \bibinfo{year}{2017}\natexlab{}.
\newblock \showarticletitle{{NEXTLEAP: Decentralizing Identity with Privacy for
  Secure Messaging}}. In \bibinfo{booktitle}{\emph{Proceedings of the 12th
  International Conference on Availability, Reliability and Security}} (Reggio
  Calabria, Italy) \emph{(\bibinfo{series}{ARES '17})}.
  \bibinfo{publisher}{ACM}, \bibinfo{address}{New York, NY, USA},
  \bibinfo{pages}{92:1--92:10}.
\newblock
\showISBNx{978-1-4503-5257-4}
\urldef\tempurl%
\url{https://doi.org/10.1145/3098954.3104056}
\showDOI{\tempurl}


\bibitem[\protect\citeauthoryear{Hardt}{Hardt}{2012}]%
        {RFC6749}
\bibfield{author}{\bibinfo{person}{D. Hardt}.} \bibinfo{year}{2012}\natexlab{}.
\newblock \bibinfo{booktitle}{\emph{The OAuth 2.0 Authorization Framework}}.
\newblock \bibinfo{type}{RFC} 6749. \bibinfo{institution}{RFC Editor}.
\newblock
\showISSN{2070-1721}
\urldef\tempurl%
\url{http://www.rfc-editor.org/rfc/rfc6749.txt}
\showURL{%
\tempurl}
\newblock
\shownote{\url{http://www.rfc-editor.org/rfc/rfc6749.txt}.}


\bibitem[\protect\citeauthoryear{Hedberg, Jones, Solberg, Gulliksson, and
  Bradley}{Hedberg et~al\mbox{.}}{2019}]%
        {oidcfederation}
\bibfield{author}{\bibinfo{person}{Roland Hedberg}, \bibinfo{person}{Michael~B.
  Jones}, \bibinfo{person}{Andreas Solberg}, \bibinfo{person}{Samuel
  Gulliksson}, {and} \bibinfo{person}{John Bradley}.}
  \bibinfo{year}{2019}\natexlab{}.
\newblock \bibinfo{booktitle}{\emph{{OpenID Connect Federation 1.0 - draft
  10}}}.
\newblock \bibinfo{type}{OpenID Specification}. \bibinfo{institution}{OpenID
  Foundation}.
\newblock


\bibitem[\protect\citeauthoryear{Hernandez-Ardieta, Heppe, and
  Carvajal-Vion}{Hernandez-Ardieta et~al\mbox{.}}{2010}]%
        {5514447}
\bibfield{author}{\bibinfo{person}{Jorge~Lopez Hernandez-Ardieta},
  \bibinfo{person}{John Heppe}, {and} \bibinfo{person}{Jose~Fernando
  Carvajal-Vion}.} \bibinfo{year}{2010}\natexlab{}.
\newblock \showarticletitle{{STORK: The European Electronic Identity
  Interoperability Platform}}.
\newblock \bibinfo{journal}{\emph{{IEEE Latin America Transactions}}}
  \bibinfo{volume}{8}, \bibinfo{number}{2} (\bibinfo{date}{April}
  \bibinfo{year}{2010}), \bibinfo{pages}{190--193}.
\newblock
\showISSN{1548-0992}
\urldef\tempurl%
\url{https://doi.org/10.1109/TLA.2010.5514447}
\showDOI{\tempurl}


\bibitem[\protect\citeauthoryear{{H{\"o}randner}, {Krenn}, {Migliavacca},
  {Thiemer}, and {Zwattendorfer}}{{H{\"o}randner} et~al\mbox{.}}{2016}]%
        {7784641}
\bibfield{author}{\bibinfo{person}{Felix {H{\"o}randner}},
  \bibinfo{person}{Stephan {Krenn}}, \bibinfo{person}{Andrea {Migliavacca}},
  \bibinfo{person}{Florian {Thiemer}}, {and} \bibinfo{person}{Bernd
  {Zwattendorfer}}.} \bibinfo{year}{2016}\natexlab{}.
\newblock \showarticletitle{{CREDENTIAL: A Framework for Privacy-Preserving
  Cloud-Based Data Sharing}}. In \bibinfo{booktitle}{\emph{2016 11th
  International Conference on Availability, Reliability and Security (ARES)}}.
  \bibinfo{pages}{742--749}.
\newblock
\urldef\tempurl%
\url{https://doi.org/10.1109/ARES.2016.79}
\showDOI{\tempurl}


\bibitem[\protect\citeauthoryear{Horrow and Sardana}{Horrow and
  Sardana}{2012}]%
        {10.1145/2490428.2490456}
\bibfield{author}{\bibinfo{person}{Susmita Horrow} {and}
  \bibinfo{person}{Anjali Sardana}.} \bibinfo{year}{2012}\natexlab{}.
\newblock \showarticletitle{{Identity Management Framework for Cloud Based
  Internet of Things}}. In \bibinfo{booktitle}{\emph{{Proceedings of the First
  International Conference on Security of Internet of Things}}} (Kollam, India)
  \emph{(\bibinfo{series}{SecurIT 12})}. \bibinfo{publisher}{Association for
  Computing Machinery}, \bibinfo{address}{New York, NY, USA},
  \bibinfo{pages}{200--203}.
\newblock
\showISBNx{9781450318228}
\urldef\tempurl%
\url{https://doi.org/10.1145/2490428.2490456}
\showDOI{\tempurl}


\bibitem[\protect\citeauthoryear{Hulsebosch, Lenzini, and Eertink}{Hulsebosch
  et~al\mbox{.}}{2009}]%
        {stork}
\bibfield{author}{\bibinfo{person}{Bob Hulsebosch}, \bibinfo{person}{Gabriele
  Lenzini}, {and} \bibinfo{person}{Henk Eertink}.}
  \bibinfo{year}{2009}\natexlab{}.
\newblock \bibinfo{booktitle}{\emph{{D2.3 -- Quality authenticator scheme}}}.
\newblock \bibinfo{type}{Deliverable}. \bibinfo{institution}{{STORK-eID
  Consortium}}.
\newblock


\bibitem[\protect\citeauthoryear{{Hulsebosch}, {Bargh}, {Fennema}, {Zandbelt},
  {Snijders}, and {Eertink}}{{Hulsebosch} et~al\mbox{.}}{2006}]%
        {1690550}
\bibfield{author}{\bibinfo{person}{R.~J. {Hulsebosch}},
  \bibinfo{person}{Mortaza {Bargh}}, \bibinfo{person}{P.~H. {Fennema}},
  \bibinfo{person}{J.~F. {Zandbelt}}, \bibinfo{person}{Martin {Snijders}},
  {and} \bibinfo{person}{Henk {Eertink}}.} \bibinfo{year}{2006}\natexlab{}.
\newblock \showarticletitle{{Using Identity Management and Secure DNS for
  Effective and Trusted User Controlled Light-Path Establishment}}. In
  \bibinfo{booktitle}{\emph{International conference on Networking and Services
  (ICNS'06)}}. \bibinfo{pages}{79--79}.
\newblock
\urldef\tempurl%
\url{https://doi.org/10.1109/ICNS.2006.115}
\showDOI{\tempurl}


\bibitem[\protect\citeauthoryear{Isaakidis, Halpin, and Danezis}{Isaakidis
  et~al\mbox{.}}{2016}]%
        {Isaakidis:2016:UPF:2994620.2994637}
\bibfield{author}{\bibinfo{person}{Marios Isaakidis}, \bibinfo{person}{Harry
  Halpin}, {and} \bibinfo{person}{George Danezis}.}
  \bibinfo{year}{2016}\natexlab{}.
\newblock \showarticletitle{{UnlimitID: Privacy-Preserving Federated Identity
  Management Using Algebraic MACs}}. In \bibinfo{booktitle}{\emph{Proceedings
  of the 2016 ACM on Workshop on Privacy in the Electronic Society}} (Vienna,
  Austria) \emph{(\bibinfo{series}{WPES '16})}. \bibinfo{publisher}{ACM},
  \bibinfo{address}{New York, NY, USA}, \bibinfo{pages}{139--142}.
\newblock
\showISBNx{978-1-4503-4569-9}
\urldef\tempurl%
\url{https://doi.org/10.1145/2994620.2994637}
\showDOI{\tempurl}


\bibitem[\protect\citeauthoryear{Jesus, Izquierdo-Moreno, Vasile-Cabezas, and
  Garcia-Blas}{Jesus et~al\mbox{.}}{2018}]%
        {8543142}
\bibfield{author}{\bibinfo{person}{Carretero Jesus}, \bibinfo{person}{Guillermo
  Izquierdo-Moreno}, \bibinfo{person}{Mario Vasile-Cabezas}, {and}
  \bibinfo{person}{Javier Garcia-Blas}.} \bibinfo{year}{2018}\natexlab{}.
\newblock \showarticletitle{{Federated Identity Architecture of the European
  eID System}}.
\newblock \bibinfo{journal}{\emph{IEEE Access}}  \bibinfo{volume}{6}
  (\bibinfo{year}{2018}), \bibinfo{pages}{75302--75326}.
\newblock
\showISSN{2169-3536}
\urldef\tempurl%
\url{https://doi.org/10.1109/ACCESS.2018.2882870}
\showDOI{\tempurl}


\bibitem[\protect\citeauthoryear{Jeyakumar, Wagner, and
  {Ro{\ss}nagel}}{Jeyakumar et~al\mbox{.}}{2019}]%
        {mci/Jeyakumar2019}
\bibfield{author}{\bibinfo{person}{Isaac Henderson~Johnson Jeyakumar},
  \bibinfo{person}{Sven Wagner}, {and} \bibinfo{person}{Heiko {Ro{\ss}nagel}}.}
  \bibinfo{year}{2019}\natexlab{}.
\newblock \showarticletitle{{Implementation of Distributed Light weight trust
  infrastructure for automatic validation of faults in an IOT sensor network}}.
  In \bibinfo{booktitle}{\emph{Open Identity Summit 2019}},
  \bibfield{editor}{\bibinfo{person}{Heiko {Ro{\ss}nagel}},
  \bibinfo{person}{Sven Wagner}, {and} \bibinfo{person}{Detlef {H{\"u}hnlein}}}
  (Eds.). \bibinfo{publisher}{Gesellschaft f{\"u}r Informatik, Bonn},
  \bibinfo{pages}{83--93}.
\newblock


\bibitem[\protect\citeauthoryear{Johansson}{Johansson}{2012}]%
        {iana}
\bibfield{author}{\bibinfo{person}{Leif Johansson}.}
  \bibinfo{year}{2012}\natexlab{}.
\newblock \bibinfo{booktitle}{\emph{{An IANA Registry for Level of Assurance
  (LoA) Profiles}}}.
\newblock \bibinfo{type}{RFC} 6711. \bibinfo{institution}{RFC Editor}.
\newblock
\showISSN{2070-1721}


\bibitem[\protect\citeauthoryear{{Kang} and {Khashnobish}}{{Kang} and
  {Khashnobish}}{2009}]%
        {5070648}
\bibfield{author}{\bibinfo{person}{Myong {Kang}} {and} \bibinfo{person}{Amitabh
  {Khashnobish}}.} \bibinfo{year}{2009}\natexlab{}.
\newblock \showarticletitle{{A Peer-to-Peer Federated Authentication System}}.
  In \bibinfo{booktitle}{\emph{2009 Sixth International Conference on
  Information Technology: New Generations}}. \bibinfo{pages}{382--387}.
\newblock
\urldef\tempurl%
\url{https://doi.org/10.1109/ITNG.2009.159}
\showDOI{\tempurl}


\bibitem[\protect\citeauthoryear{{Kantara Initiative}}{{Kantara
  Initiative}}{2015}]%
        {uma}
\bibfield{author}{\bibinfo{person}{{Kantara Initiative}}.}
  \bibinfo{year}{2015}\natexlab{}.
\newblock \bibinfo{title}{{Home -- WG -- User Managed Access}}.
\newblock
  \bibinfo{howpublished}{\url{https://kantarainitiative.org/confluence/display/uma/Home}}.
\newblock
\newblock
\shownote{[Online, \today].}


\bibitem[\protect\citeauthoryear{Karegar, Striecks, Krenn, H{\"o}randner,
  Lor{\"u}nser, and Fischer-H{\"u}bner}{Karegar et~al\mbox{.}}{2016}]%
        {Karegar2016}
\bibfield{author}{\bibinfo{person}{Farzaneh Karegar},
  \bibinfo{person}{Christoph Striecks}, \bibinfo{person}{Stephan Krenn},
  \bibinfo{person}{Felix H{\"o}randner}, \bibinfo{person}{Thomas Lor{\"u}nser},
  {and} \bibinfo{person}{Simone Fischer-H{\"u}bner}.}
  \bibinfo{year}{2016}\natexlab{}.
\newblock \bibinfo{booktitle}{\emph{{Opportunities and Challenges of
  CREDENTIAL}}}.
\newblock \bibinfo{publisher}{Springer International Publishing},
  \bibinfo{address}{Cham}, \bibinfo{pages}{76--91}.
\newblock
\showISBNx{978-3-319-55783-0}
\urldef\tempurl%
\url{https://doi.org/10.1007/978-3-319-55783-0_7}
\showDOI{\tempurl}


\bibitem[\protect\citeauthoryear{Keltoum and Samia}{Keltoum and Samia}{2017}]%
        {Keltoum:2017:DFI:3018896.3025152}
\bibfield{author}{\bibinfo{person}{Bendiab Keltoum} {and}
  \bibinfo{person}{Boucherkha Samia}.} \bibinfo{year}{2017}\natexlab{}.
\newblock \showarticletitle{{A Dynamic Federated Identity Management Approach
  for Cloud-based Environments}}. In \bibinfo{booktitle}{\emph{Proceedings of
  the Second International Conference on Internet of Things, Data and Cloud
  Computing}} (Cambridge, United Kingdom) \emph{(\bibinfo{series}{ICC '17})}.
  \bibinfo{publisher}{ACM}, \bibinfo{address}{New York, NY, USA}, Article
  \bibinfo{articleno}{104}, \bibinfo{numpages}{5}~pages.
\newblock
\showISBNx{978-1-4503-4774-7}
\urldef\tempurl%
\url{https://doi.org/10.1145/3018896.3025152}
\showDOI{\tempurl}


\bibitem[\protect\citeauthoryear{{Kobayashi} and {Talburt}}{{Kobayashi} and
  {Talburt}}{2014}]%
        {6822222}
\bibfield{author}{\bibinfo{person}{Fumiko {Kobayashi}} {and}
  \bibinfo{person}{John~R. {Talburt}}.} \bibinfo{year}{2014}\natexlab{}.
\newblock \showarticletitle{{Decoupling Identity Resolution from the
  Maintenance of Identity Information}}. In \bibinfo{booktitle}{\emph{2014 11th
  International Conference on Information Technology: New Generations}}.
  \bibinfo{pages}{349--354}.
\newblock
\urldef\tempurl%
\url{https://doi.org/10.1109/ITNG.2014.88}
\showDOI{\tempurl}


\bibitem[\protect\citeauthoryear{Kostopoulos, Sfakianakis, Chochliouros,
  Pettersson, Krenn, Tesfay, Migliavacca, and H\"{o}randner}{Kostopoulos
  et~al\mbox{.}}{2017}]%
        {Kostopoulos:2017:TAS:3098954.3104061}
\bibfield{author}{\bibinfo{person}{Alexandros Kostopoulos},
  \bibinfo{person}{Evangelos Sfakianakis}, \bibinfo{person}{Ioannis
  Chochliouros}, \bibinfo{person}{John~S\"{o}ren Pettersson},
  \bibinfo{person}{Stephan Krenn}, \bibinfo{person}{Welderufael Tesfay},
  \bibinfo{person}{Andrea Migliavacca}, {and} \bibinfo{person}{Felix
  H\"{o}randner}.} \bibinfo{year}{2017}\natexlab{}.
\newblock \showarticletitle{{Towards the Adoption of Secure Cloud Identity
  Services}}. In \bibinfo{booktitle}{\emph{Proceedings of the 12th
  International Conference on Availability, Reliability and Security}} (Reggio
  Calabria, Italy) \emph{(\bibinfo{series}{ARES '17})}.
  \bibinfo{publisher}{ACM}, \bibinfo{address}{New York, NY, USA},
  \bibinfo{pages}{90:1--90:7}.
\newblock
\showISBNx{978-1-4503-5257-4}
\urldef\tempurl%
\url{https://doi.org/10.1145/3098954.3104061}
\showDOI{\tempurl}


\bibitem[\protect\citeauthoryear{Kraft}{Kraft}{2016}]%
        {nameid}
\bibfield{author}{\bibinfo{person}{Daniel Kraft}.}
  \bibinfo{year}{2016}\natexlab{}.
\newblock \bibinfo{title}{{Namecoin + OpenID = NameID!}}
\newblock \bibinfo{howpublished}{\url{https://nameid.org}}.
\newblock
\newblock
\shownote{[Online, \today].}


\bibitem[\protect\citeauthoryear{{Kumar Srivastava}, {Roychoudhury}, and
  {Vardhan Samalia}}{{Kumar Srivastava} et~al\mbox{.}}{2018}]%
        {8442455}
\bibfield{author}{\bibinfo{person}{Deepesh {Kumar Srivastava}},
  \bibinfo{person}{Basav {Roychoudhury}}, {and} \bibinfo{person}{Harsh {Vardhan
  Samalia}}.} \bibinfo{year}{2018}\natexlab{}.
\newblock \showarticletitle{{Importance of User's Profile Attributes in
  Identity Matching Across Multiple Online Social Networking Sites}}. In
  \bibinfo{booktitle}{\emph{2018 8th International Conference on Cloud
  Computing, Data Science Engineering (Confluence)}}. \bibinfo{pages}{14--15}.
\newblock
\urldef\tempurl%
\url{https://doi.org/10.1109/CONFLUENCE.2018.8442455}
\showDOI{\tempurl}


\bibitem[\protect\citeauthoryear{{Lee}, {Jeong}, and {Park}}{{Lee}
  et~al\mbox{.}}{2016}]%
        {7423411}
\bibfield{author}{\bibinfo{person}{Sejun {Lee}}, \bibinfo{person}{Jaehoon~Paul
  {Jeong}}, {and} \bibinfo{person}{Jung-Soo {Park}}.}
  \bibinfo{year}{2016}\natexlab{}.
\newblock \showarticletitle{{DNSNA: DNS name autoconfiguration for Internet of
  Things devices}}. In \bibinfo{booktitle}{\emph{2016 18th International
  Conference on Advanced Communication Technology (ICACT)}}.
  \bibinfo{pages}{1--1}.
\newblock
\urldef\tempurl%
\url{https://doi.org/10.1109/ICACT.2016.7423411}
\showDOI{\tempurl}


\bibitem[\protect\citeauthoryear{{Lu}, {Yeh}, and {Huang}}{{Lu}
  et~al\mbox{.}}{2018}]%
        {8422733}
\bibfield{author}{\bibinfo{person}{Peggy~Joy {Lu}}, \bibinfo{person}{Lo-Yao
  {Yeh}}, {and} \bibinfo{person}{Jiun-Long {Huang}}.}
  \bibinfo{year}{2018}\natexlab{}.
\newblock \showarticletitle{{An Privacy-Preserving Cross-Organizational
  Authentication/Authorization/Accounting System Using Blockchain Technology}}.
  In \bibinfo{booktitle}{\emph{2018 IEEE International Conference on
  Communications (ICC)}}. \bibinfo{pages}{1--6}.
\newblock
\showISSN{1938--1883}
\urldef\tempurl%
\url{https://doi.org/10.1109/ICC.2018.8422733}
\showDOI{\tempurl}


\bibitem[\protect\citeauthoryear{Machulak, Maler, Catalano, and van
  Moorsel}{Machulak et~al\mbox{.}}{2010}]%
        {10.1145/1866855.1866865}
\bibfield{author}{\bibinfo{person}{Maciej~P. Machulak}, \bibinfo{person}{Eve~L.
  Maler}, \bibinfo{person}{Domenico Catalano}, {and} \bibinfo{person}{Aad van
  Moorsel}.} \bibinfo{year}{2010}\natexlab{}.
\newblock \showarticletitle{{User-Managed Access to Web Resources}}. In
  \bibinfo{booktitle}{\emph{{Proceedings of the 6th ACM Workshop on Digital
  Identity Management}}} (Chicago, Illinois, USA) \emph{(\bibinfo{series}{DIM
  10})}. \bibinfo{publisher}{{Association for Computing Machinery}},
  \bibinfo{address}{New York, NY, USA}, \bibinfo{pages}{35--44}.
\newblock
\showISBNx{9781450300902}
\urldef\tempurl%
\url{https://doi.org/10.1145/1866855.1866865}
\showDOI{\tempurl}


\bibitem[\protect\citeauthoryear{Mahalle}{Mahalle}{2014}]%
        {04edd2148d064f3fa0e2addc0d6da508}
\bibfield{author}{\bibinfo{person}{{Parikshit N.} Mahalle}.}
  \bibinfo{year}{2014}\natexlab{}.
\newblock \emph{\bibinfo{title}{{Identity Management Framework for Internet of
  Things}}}.
\newblock Dissertation.
\newblock
\showISBNx{978-87-7152.031-6}


\bibitem[\protect\citeauthoryear{{Maler}}{{Maler}}{2015}]%
        {7163222}
\bibfield{author}{\bibinfo{person}{Eve {Maler}}.}
  \bibinfo{year}{2015}\natexlab{}.
\newblock \showarticletitle{{Extending the Power of Consent with User-Managed
  Access: A Standard Architecture for Asynchronous, Centralizable,
  Internet-Scalable Consent}}. In \bibinfo{booktitle}{\emph{2015 IEEE Security
  and Privacy Workshops}}. \bibinfo{pages}{175--179}.
\newblock
\urldef\tempurl%
\url{https://doi.org/10.1109/SPW.2015.34}
\showDOI{\tempurl}


\bibitem[\protect\citeauthoryear{{Moreno}, {Bernabe}, {Skarmeta}, {Stausholm},
  {Frederiksen}, {Martínez}, {Ponte}, {Sakkopoulos}, and {Lehmann}}{{Moreno}
  et~al\mbox{.}}{2019}]%
        {8766357}
\bibfield{author}{\bibinfo{person}{Rafael~Torres {Moreno}},
  \bibinfo{person}{Jorge~Bernal {Bernabe}}, \bibinfo{person}{Antonio
  {Skarmeta}}, \bibinfo{person}{Michael {Stausholm}},
  \bibinfo{person}{Tore~Kasper {Frederiksen}}, \bibinfo{person}{Noelia
  {Martínez}}, \bibinfo{person}{Nuno {Ponte}}, \bibinfo{person}{Evangelos
  {Sakkopoulos}}, {and} \bibinfo{person}{Anja {Lehmann}}.}
  \bibinfo{year}{2019}\natexlab{}.
\newblock \showarticletitle{{OLYMPUS: towards Oblivious identitY Management for
  Private and User-friendly Services}}. In \bibinfo{booktitle}{\emph{2019
  Global IoT Summit (GIoTS)}}. \bibinfo{pages}{1--6}.
\newblock
\urldef\tempurl%
\url{https://doi.org/10.1109/GIOTS.2019.8766357}
\showDOI{\tempurl}


\bibitem[\protect\citeauthoryear{{Oppliger}}{{Oppliger}}{2003}]%
        {1212687}
\bibfield{author}{\bibinfo{person}{R. {Oppliger}}.}
  \bibinfo{year}{2003}\natexlab{}.
\newblock \showarticletitle{Microsoft .Net Passport: a security analysis}.
\newblock \bibinfo{journal}{\emph{Computer}} \bibinfo{volume}{36},
  \bibinfo{number}{7} (\bibinfo{date}{July} \bibinfo{year}{2003}),
  \bibinfo{pages}{29--35}.
\newblock
\showISSN{1558-0814}
\urldef\tempurl%
\url{https://doi.org/10.1109/MC.2003.1212687}
\showDOI{\tempurl}


\bibitem[\protect\citeauthoryear{{Othman} and {Callahan}}{{Othman} and
  {Callahan}}{2018}]%
        {8489316}
\bibfield{author}{\bibinfo{person}{Asem {Othman}} {and} \bibinfo{person}{John
  {Callahan}}.} \bibinfo{year}{2018}\natexlab{}.
\newblock \showarticletitle{{The Horcrux Protocol: A Method for Decentralized
  Biometric-based Self-sovereign Identity}}. In \bibinfo{booktitle}{\emph{2018
  International Joint Conference on Neural Networks (IJCNN)}}.
  \bibinfo{pages}{1--7}.
\newblock
\showISSN{2161-4407}
\urldef\tempurl%
\url{https://doi.org/10.1109/IJCNN.2018.8489316}
\showDOI{\tempurl}


\bibitem[\protect\citeauthoryear{Othman and Callahan}{Othman and
  Callahan}{2019}]%
        {Othman2019}
\bibfield{author}{\bibinfo{person}{Asem Othman} {and} \bibinfo{person}{John
  Callahan}.} \bibinfo{year}{2019}\natexlab{}.
\newblock \bibinfo{booktitle}{\emph{The Horcrux Protocol: A Distributed Mobile
  Biometric Self-sovereign Identity Protocol}}.
\newblock \bibinfo{publisher}{Springer International Publishing},
  \bibinfo{address}{Cham}, \bibinfo{pages}{355--377}.
\newblock
\showISBNx{978-3-030-26972-2}
\urldef\tempurl%
\url{https://doi.org/10.1007/978-3-030-26972-2_17}
\showDOI{\tempurl}


\bibitem[\protect\citeauthoryear{P{\"o}hn, Metzger, and Hommel}{P{\"o}hn
  et~al\mbox{.}}{2014a}]%
        {infocomp2014}
\bibfield{author}{\bibinfo{person}{Daniela P{\"o}hn}, \bibinfo{person}{Stefan
  Metzger}, {and} \bibinfo{person}{Wolfgang Hommel}.}
  \bibinfo{year}{2014}\natexlab{a}.
\newblock \showarticletitle{{A SAML Metadata Broker for Dynamic Federations and
  Inter-Federations}}. In \bibinfo{booktitle}{\emph{Proceedings of INFOCOMP
  2014, The Fourth International Conference on Advanced Communications and
  Computation}} (Paris, France). \bibinfo{publisher}{IARIA},
  \bibinfo{pages}{132--137}.
\newblock
\newblock
\shownote{{ISBN:} 978-1-61208-365-0.}


\bibitem[\protect\citeauthoryear{P{\"o}hn, Metzger, and Hommel}{P{\"o}hn
  et~al\mbox{.}}{2014b}]%
        {ifipsec2014}
\bibfield{author}{\bibinfo{person}{Daniela P{\"o}hn}, \bibinfo{person}{Stefan
  Metzger}, {and} \bibinfo{person}{Wolfgang Hommel}.}
  \bibinfo{year}{2014}\natexlab{b}.
\newblock \showarticletitle{{G{\'e}ant-TrustBroker: Dynamic, Scalable
  Management of SAML-Based Inter-federation Authentication and Authorization
  Infrastructures}}.
\newblock In \bibinfo{booktitle}{\emph{ICT Systems Security and Privacy
  Protection}}, \bibfield{editor}{\bibinfo{person}{Nora Cuppens-Boulahia},
  \bibinfo{person}{Fr{\'e}d{\'e}ric Cuppens}, \bibinfo{person}{Sushil Jajodia},
  \bibinfo{person}{Anas Abou~El~Kalam}, {and} \bibinfo{person}{Thierry Sans}}
  (Eds.). \bibinfo{series}{IFIP Advances in Information and Communication
  Technology}, Vol.~\bibinfo{volume}{428}. \bibinfo{publisher}{Springer Berlin
  Heidelberg}, \bibinfo{pages}{307--320}.
\newblock
\urldef\tempurl%
\url{https://doi.org/10.1007/978-3-642-55415-5_25}
\showDOI{\tempurl}


\bibitem[\protect\citeauthoryear{Ragouzis, Hughes, Philpott, and
  Maler}{Ragouzis et~al\mbox{.}}{2008}]%
        {samloverview}
\bibfield{author}{\bibinfo{person}{Nick Ragouzis}, \bibinfo{person}{John
  Hughes}, \bibinfo{person}{Rob Philpott}, {and} \bibinfo{person}{Eve Maler}.}
  \bibinfo{year}{2008}\natexlab{}.
\newblock \bibinfo{booktitle}{\emph{{Security Assertion Markup Language (SAML)
  V2.0 Technical Overview}}}.
\newblock \bibinfo{type}{{T}echnical {R}eport}. \bibinfo{institution}{OASIS}.
\newblock


\bibitem[\protect\citeauthoryear{Ribeiro, Leitold, Esposito, and
  Mitzam}{Ribeiro et~al\mbox{.}}{2018}]%
        {Ribeiro:2018:SRH:3280301.3280330}
\bibfield{author}{\bibinfo{person}{Carlos Ribeiro}, \bibinfo{person}{Herbert
  Leitold}, \bibinfo{person}{Simon Esposito}, {and} \bibinfo{person}{David
  Mitzam}.} \bibinfo{year}{2018}\natexlab{}.
\newblock \showarticletitle{{STORK: A Real, Heterogeneous, Large-scale eID
  Management System}}.
\newblock \bibinfo{journal}{\emph{Int. J. Inf. Secur.}} \bibinfo{volume}{17},
  \bibinfo{number}{5} (\bibinfo{date}{Oct.} \bibinfo{year}{2018}),
  \bibinfo{pages}{569--585}.
\newblock
\showISSN{1615-5262}
\urldef\tempurl%
\url{https://doi.org/10.1007/s10207-017-0385-x}
\showDOI{\tempurl}


\bibitem[\protect\citeauthoryear{Richer and Johansson}{Richer and
  Johansson}{2018}]%
        {RFC8485}
\bibfield{author}{\bibinfo{person}{J. Richer} {and} \bibinfo{person}{L.
  Johansson}.} \bibinfo{year}{2018}\natexlab{}.
\newblock \bibinfo{booktitle}{\emph{Vectors of Trust}}.
\newblock \bibinfo{type}{RFC} 8485. \bibinfo{institution}{RFC Editor}.
\newblock
\showISSN{2070-1721}
\urldef\tempurl%
\url{http://www.rfc-editor.org/rfc/rfc8485.txt}
\showURL{%
\tempurl}
\newblock
\shownote{\url{http://www.rfc-editor.org/rfc/rfc8485.txt}.}


\bibitem[\protect\citeauthoryear{Ro{\ss}nagel}{Ro{\ss}nagel}{2017}]%
        {mci/Rossnagel2017}
\bibfield{author}{\bibinfo{person}{Heiko Ro{\ss}nagel}.}
  \bibinfo{year}{2017}\natexlab{}.
\newblock \showarticletitle{{A Mechanism for Discovery and Verification of
  Trust Scheme Memberships: The Lightest Reference Architecture}}. In
  \bibinfo{booktitle}{\emph{Open Identity Summit 2017}},
  \bibfield{editor}{\bibinfo{person}{Lothar Fritsch}, \bibinfo{person}{Heiko
  {Ro{\ss}nagel}}, {and} \bibinfo{person}{Detlef {H{\"u}hnlein}}} (Eds.).
  \bibinfo{publisher}{Gesellschaft f{\"u}r Informatik, Bonn},
  \bibinfo{pages}{81--92}.
\newblock


\bibitem[\protect\citeauthoryear{Ro{\ss}nagel and Wagner}{Ro{\ss}nagel and
  Wagner}{2019}]%
        {Rossnagel2019}
\bibfield{author}{\bibinfo{person}{Heiko Ro{\ss}nagel} {and}
  \bibinfo{person}{Sven Wagner}.} \bibinfo{year}{2019}\natexlab{}.
\newblock \showarticletitle{{LIGHTest}}.
\newblock \bibinfo{journal}{\emph{{Datenschutz und Datensicherheit - DuD}}}
  \bibinfo{volume}{43}, \bibinfo{number}{4} (\bibinfo{date}{Apr}
  \bibinfo{year}{2019}), \bibinfo{pages}{220--224}.
\newblock
\showISSN{1862-2607}
\urldef\tempurl%
\url{https://doi.org/10.1007/s11623-019-1096-4}
\showDOI{\tempurl}


\bibitem[\protect\citeauthoryear{Sakimura, Bradley, Jones, de~Medeiros, and
  Mortimore}{Sakimura et~al\mbox{.}}{2014}]%
        {openidconnect}
\bibfield{author}{\bibinfo{person}{Nat Sakimura}, \bibinfo{person}{John
  Bradley}, \bibinfo{person}{Michael~B. Jones}, \bibinfo{person}{Breno de
  Medeiros}, {and} \bibinfo{person}{Chuck Mortimore}.}
  \bibinfo{year}{2014}\natexlab{}.
\newblock \bibinfo{booktitle}{\emph{{OpenID Connect Core 1.0}}}.
\newblock \bibinfo{type}{{T}echnical {R}eport}. \bibinfo{institution}{OpenID
  Foundation}.
\newblock


\bibitem[\protect\citeauthoryear{Schanzenbach and Banse}{Schanzenbach and
  Banse}{2016}]%
        {schanzenbach2016}
\bibfield{author}{\bibinfo{person}{Martin Schanzenbach} {and}
  \bibinfo{person}{Christian Banse}.} \bibinfo{year}{2016}\natexlab{}.
\newblock \showarticletitle{{Managing and Presenting User Attributes over a
  Decentralized Secure Name System}}. In \bibinfo{booktitle}{\emph{Data privacy
  management and security assurance. 11th International Workshop, DPM 2016 and
  5th International Workshop, QASA 2016}}. \bibinfo{publisher}{European
  Symposium on Research in Computer Security}, \bibinfo{address}{Heraklion,
  Crete, Greece}, \bibinfo{pages}{213--220}.
\newblock
\urldef\tempurl%
\url{http://dx.doi.org/10.1007/978-3-319-47072-6_14}
\showURL{%
\tempurl}


\bibitem[\protect\citeauthoryear{Schanzenbach, Banse, and
  Sch{\"u}tte}{Schanzenbach et~al\mbox{.}}{2018}]%
        {schanzenbach2018abd}
\bibfield{author}{\bibinfo{person}{Martin Schanzenbach},
  \bibinfo{person}{Christian Banse}, {and} \bibinfo{person}{Julian
  Sch{\"u}tte}.} \bibinfo{year}{2018}\natexlab{}.
\newblock \showarticletitle{{ Practical Decentralized Attribute-Based
  Delegation Using Secure Name Systems }}. In
  \bibinfo{booktitle}{\emph{International Conference on Trust, Security and
  Privacy in Computing and Communications (TrustCom)}}.
  \bibinfo{publisher}{IEEE}, \bibinfo{pages}{244--251}.
\newblock
\urldef\tempurl%
\url{http://dx.doi.org/10.1109/TrustCom/BigDataSE.2018.00046}
\showURL{%
\tempurl}


\bibitem[\protect\citeauthoryear{{Schanzenbach}, {Bramm}, and
  {Sch{\"u}tte}}{{Schanzenbach} et~al\mbox{.}}{2018}]%
        {8456003}
\bibfield{author}{\bibinfo{person}{Martin {Schanzenbach}},
  \bibinfo{person}{Georg {Bramm}}, {and} \bibinfo{person}{Julian
  {Sch{\"u}tte}}.} \bibinfo{year}{2018}\natexlab{}.
\newblock \showarticletitle{{reclaimID: Secure, Self-Sovereign Identities Using
  Name Systems and Attribute-Based Encryption}}. In
  \bibinfo{booktitle}{\emph{2018 17th IEEE International Conference On Trust,
  Security And Privacy In Computing And Communications/ 12th IEEE International
  Conference On Big Data Science And Engineering (TrustCom/BigDataSE)}}.
  \bibinfo{pages}{946--957}.
\newblock
\showISSN{2324-9013}
\urldef\tempurl%
\url{https://doi.org/10.1109/TrustCom/BigDataSE.2018.00134}
\showDOI{\tempurl}


\bibitem[\protect\citeauthoryear{Schwartz}{Schwartz}{2013}]%
        {samluma}
\bibfield{author}{\bibinfo{person}{Michael Schwartz}.}
  \bibinfo{year}{2013}\natexlab{}.
\newblock \bibinfo{title}{{Recipe for a Reverse Proxy using SAML and UMA}}.
\newblock
  \bibinfo{howpublished}{\url{https://www.gluu.org/blog/recipe-for-a-reverse-proxy-using-saml-and-uma/}}.
\newblock
\newblock
\shownote{[Online, \today].}


\bibitem[\protect\citeauthoryear{Seitz, Gerdes, Selander, Mani, and
  Kumar}{Seitz et~al\mbox{.}}{2016}]%
        {RFC7744}
\bibfield{author}{\bibinfo{person}{L. Seitz}, \bibinfo{person}{S. Gerdes},
  \bibinfo{person}{G. Selander}, \bibinfo{person}{M. Mani}, {and}
  \bibinfo{person}{S. Kumar}.} \bibinfo{year}{2016}\natexlab{}.
\newblock \bibinfo{booktitle}{\emph{{Use Cases for Authentication and
  Authorization in Constrained Environments}}}.
\newblock \bibinfo{type}{RFC} 7744. \bibinfo{institution}{RFC Editor}.
\newblock
\showISSN{2070-1721}


\bibitem[\protect\citeauthoryear{Slomovic}{Slomovic}{2014}]%
        {6837386}
\bibfield{author}{\bibinfo{person}{Anna Slomovic}.}
  \bibinfo{year}{2014}\natexlab{}.
\newblock \showarticletitle{{Privacy Issues in Identity Verification}}.
\newblock \bibinfo{journal}{\emph{IEEE Security Privacy}} \bibinfo{volume}{12},
  \bibinfo{number}{3} (\bibinfo{date}{May} \bibinfo{year}{2014}),
  \bibinfo{pages}{71--73}.
\newblock
\showISSN{1540-7993}
\urldef\tempurl%
\url{https://doi.org/10.1109/MSP.2014.52}
\showDOI{\tempurl}


\bibitem[\protect\citeauthoryear{Speck}{Speck}{2019}]%
        {id4me}
\bibfield{author}{\bibinfo{person}{Katja Speck}.}
  \bibinfo{year}{2019}\natexlab{}.
\newblock \bibinfo{title}{{Independent, Federated Digital Identity Management
  Solution ID4me Announces Public Beta at CloudFest 2019}}.
\newblock
  \bibinfo{howpublished}{\url{https://id4me.org/independent-federated-digital-identity-management-solution\\-id4me-announces-public-beta-at-cloudfest-2019/}}.
\newblock
\newblock
\shownote{[Online, \today].}


\bibitem[\protect\citeauthoryear{{Steichen}, {Fiz}, {Norvill}, {Shbair}, and
  {State}}{{Steichen} et~al\mbox{.}}{2018}]%
        {8726493}
\bibfield{author}{\bibinfo{person}{Mathis {Steichen}}, \bibinfo{person}{Beltran
  {Fiz}}, \bibinfo{person}{Robert {Norvill}}, \bibinfo{person}{Wazen {Shbair}},
  {and} \bibinfo{person}{Radu {State}}.} \bibinfo{year}{2018}\natexlab{}.
\newblock \showarticletitle{{Blockchain-Based, Decentralized Access Control for
  IPFS}}. In \bibinfo{booktitle}{\emph{2018 IEEE International Conference on
  Internet of Things (iThings) and IEEE Green Computing and Communications
  (GreenCom) and IEEE Cyber, Physical and Social Computing (CPSCom) and IEEE
  Smart Data (SmartData)}}. \bibinfo{pages}{1499--1506}.
\newblock
\urldef\tempurl%
\url{https://doi.org/10.1109/Cybermatics_2018.2018.00253}
\showDOI{\tempurl}


\bibitem[\protect\citeauthoryear{{Stokkink} and {Pouwelse}}{{Stokkink} and
  {Pouwelse}}{2018}]%
        {8726562}
\bibfield{author}{\bibinfo{person}{Quinten {Stokkink}} {and}
  \bibinfo{person}{Johan {Pouwelse}}.} \bibinfo{year}{2018}\natexlab{}.
\newblock \showarticletitle{{Deployment of a Blockchain-Based Self-Sovereign
  Identity}}. In \bibinfo{booktitle}{\emph{2018 IEEE International Conference
  on Internet of Things (iThings) and IEEE Green Computing and Communications
  (GreenCom) and IEEE Cyber, Physical and Social Computing (CPSCom) and IEEE
  Smart Data (SmartData)}}. \bibinfo{pages}{1336--1342}.
\newblock
\urldef\tempurl%
\url{https://doi.org/10.1109/Cybermatics_2018.2018.00230}
\showDOI{\tempurl}


\bibitem[\protect\citeauthoryear{{Takemiya} and {Vanieiev}}{{Takemiya} and
  {Vanieiev}}{2018}]%
        {8377927}
\bibfield{author}{\bibinfo{person}{Makoto {Takemiya}} {and}
  \bibinfo{person}{Bohdan {Vanieiev}}.} \bibinfo{year}{2018}\natexlab{}.
\newblock \showarticletitle{{Sora Identity: Secure, Digital Identity on the
  Blockchain}}. In \bibinfo{booktitle}{\emph{2018 IEEE 42nd Annual Computer
  Software and Applications Conference (COMPSAC)}}, Vol.~\bibinfo{volume}{02}.
  \bibinfo{pages}{582--587}.
\newblock
\showISSN{0730-3157}
\urldef\tempurl%
\url{https://doi.org/10.1109/COMPSAC.2018.10299}
\showDOI{\tempurl}


\bibitem[\protect\citeauthoryear{Tobin and Reed}{Tobin and Reed}{2017}]%
        {sovrin}
\bibfield{author}{\bibinfo{person}{Andrew Tobin} {and}
  \bibinfo{person}{Drummond Reed}.} \bibinfo{year}{2017}\natexlab{}.
\newblock \bibinfo{title}{{The Inevitable Rise of Self-Sovereign Identity}}.
\newblock
  \bibinfo{howpublished}{\url{https://sovrin.org/wp-content/uploads/2018/03/The-Inevitable-Rise-of-Self-Sovereign-Identity.pdf}}.
\newblock
\newblock
\shownote{[Online, \today].}


\bibitem[\protect\citeauthoryear{{Torres}, {Nogueira}, and {Pujolle}}{{Torres}
  et~al\mbox{.}}{2013}]%
        {6275425}
\bibfield{author}{\bibinfo{person}{J. {Torres}}, \bibinfo{person}{M.
  {Nogueira}}, {and} \bibinfo{person}{G. {Pujolle}}.}
  \bibinfo{year}{2013}\natexlab{}.
\newblock \showarticletitle{A Survey on Identity Management for the Future
  Network}.
\newblock \bibinfo{journal}{\emph{IEEE Communications Surveys Tutorials}}
  \bibinfo{volume}{15}, \bibinfo{number}{2} (\bibinfo{date}{Second}
  \bibinfo{year}{2013}), \bibinfo{pages}{787--802}.
\newblock
\showISSN{2373-745X}
\urldef\tempurl%
\url{https://doi.org/10.1109/SURV.2012.072412.00129}
\showDOI{\tempurl}


\bibitem[\protect\citeauthoryear{Toth and Anderson-Priddy}{Toth and
  Anderson-Priddy}{2019}]%
        {8713271}
\bibfield{author}{\bibinfo{person}{Kalman Toth} {and} \bibinfo{person}{Alan
  Anderson-Priddy}.} \bibinfo{year}{2019}\natexlab{}.
\newblock \showarticletitle{{Self-Sovereign Digital Identity: A Paradigm Shift
  for Identity}}.
\newblock \bibinfo{journal}{\emph{IEEE Security Privacy}} \bibinfo{volume}{17},
  \bibinfo{number}{3} (\bibinfo{date}{May} \bibinfo{year}{2019}),
  \bibinfo{pages}{17--27}.
\newblock
\showISSN{1540-7993}
\urldef\tempurl%
\url{https://doi.org/10.1109/MSEC.2018.2888782}
\showDOI{\tempurl}


\bibitem[\protect\citeauthoryear{{van Thuan}, {Butkus}, and {van Thanh}}{{van
  Thuan} et~al\mbox{.}}{2014}]%
        {7021724}
\bibfield{author}{\bibinfo{person}{Do {van Thuan}}, \bibinfo{person}{Pranas
  {Butkus}}, {and} \bibinfo{person}{Do {van Thanh}}.}
  \bibinfo{year}{2014}\natexlab{}.
\newblock \showarticletitle{{A User Centric Identity Management for Internet of
  Things}}. In \bibinfo{booktitle}{\emph{2014 International Conference on IT
  Convergence and Security (ICITCS)}}. \bibinfo{pages}{1--4}.
\newblock
\urldef\tempurl%
\url{https://doi.org/10.1109/ICITCS.2014.7021724}
\showDOI{\tempurl}


\bibitem[\protect\citeauthoryear{Veseli, Olvera, Pulls, and Rannenberg}{Veseli
  et~al\mbox{.}}{2019}]%
        {Veseli:2019:EPD:3297280.3297429}
\bibfield{author}{\bibinfo{person}{Fatbardh Veseli},
  \bibinfo{person}{Jetzabel~Serna Olvera}, \bibinfo{person}{Tobias Pulls},
  {and} \bibinfo{person}{Kai Rannenberg}.} \bibinfo{year}{2019}\natexlab{}.
\newblock \showarticletitle{{Engineering Privacy by Design: Lessons from the
  Design and Implementation of an Identity Wallet Platform}}. In
  \bibinfo{booktitle}{\emph{Proceedings of the 34th ACM/SIGAPP Symposium on
  Applied Computing}} (Limassol, Cyprus) \emph{(\bibinfo{series}{SAC '19})}.
  \bibinfo{publisher}{ACM}, \bibinfo{address}{New York, NY, USA},
  \bibinfo{pages}{1475--1483}.
\newblock
\showISBNx{978-1-4503-5933-7}
\urldef\tempurl%
\url{https://doi.org/10.1145/3297280.3297429}
\showDOI{\tempurl}


\bibitem[\protect\citeauthoryear{Wagner, Omolola, and More}{Wagner
  et~al\mbox{.}}{2017}]%
        {mci/Wagner2017}
\bibfield{author}{\bibinfo{person}{Georg Wagner}, \bibinfo{person}{Olamide
  Omolola}, {and} \bibinfo{person}{Stefan More}.}
  \bibinfo{year}{2017}\natexlab{}.
\newblock \showarticletitle{{Harmonizing Delegation Data Formats}}. In
  \bibinfo{booktitle}{\emph{Open Identity Summit 2017}},
  \bibfield{editor}{\bibinfo{person}{Lothar Fritsch}, \bibinfo{person}{Heiko
  {Ro{\ss}nagel}}, {and} \bibinfo{person}{Detlef {H{\"u}hnlein}}} (Eds.).
  \bibinfo{publisher}{Gesellschaft f{\"u}r Informatik, Bonn},
  \bibinfo{pages}{25--34}.
\newblock


\bibitem[\protect\citeauthoryear{Wagner, Wagner, More, and Hoffmann}{Wagner
  et~al\mbox{.}}{2019}]%
        {mci/Wagner2019}
\bibfield{author}{\bibinfo{person}{Georg Wagner}, \bibinfo{person}{Sven
  Wagner}, \bibinfo{person}{Stefan More}, {and} \bibinfo{person}{Martin
  Hoffmann}.} \bibinfo{year}{2019}\natexlab{}.
\newblock \showarticletitle{{DNS-based Trust Scheme Publication and
  Discovery}}. In \bibinfo{booktitle}{\emph{Open Identity Summit 2019}},
  \bibfield{editor}{\bibinfo{person}{Heiko {Ro{\ss}nagel}},
  \bibinfo{person}{Sven Wagner}, {and} \bibinfo{person}{Detlef {H{\"u}hnlein}}}
  (Eds.). \bibinfo{publisher}{Gesellschaft f{\"u}r Informatik, Bonn},
  \bibinfo{pages}{49--58}.
\newblock


\bibitem[\protect\citeauthoryear{{Yakubov}, {Shbair}, {Wallbom}, {Sanda}, and
  {State}}{{Yakubov} et~al\mbox{.}}{2018}]%
        {8406325}
\bibfield{author}{\bibinfo{person}{Alexander {Yakubov}},
  \bibinfo{person}{Wazen~M. {Shbair}}, \bibinfo{person}{Anders {Wallbom}},
  \bibinfo{person}{David {Sanda}}, {and} \bibinfo{person}{Radu {State}}.}
  \bibinfo{year}{2018}\natexlab{}.
\newblock \showarticletitle{{A blockchain-based PKI management framework}}. In
  \bibinfo{booktitle}{\emph{NOMS 2018 - 2018 IEEE/IFIP Network Operations and
  Management Symposium}}. \bibinfo{pages}{1--6}.
\newblock
\showISSN{2374-9709}
\urldef\tempurl%
\url{https://doi.org/10.1109/NOMS.2018.8406325}
\showDOI{\tempurl}


\bibitem[\protect\citeauthoryear{{Yanjiong Wang} and {Qiaoyan Wen}}{{Yanjiong
  Wang} and {Qiaoyan Wen}}{2011}]%
        {6192955}
\bibfield{author}{\bibinfo{person}{{Yanjiong Wang}} {and}
  \bibinfo{person}{{Qiaoyan Wen}}.} \bibinfo{year}{2011}\natexlab{}.
\newblock \showarticletitle{{A privacy enhanced DNS scheme for the Internet Of
  Things}}. In \bibinfo{booktitle}{\emph{IET International Conference on
  Communication Technology and Application (ICCTA 2011)}}.
  \bibinfo{pages}{699--702}.
\newblock
\urldef\tempurl%
\url{https://doi.org/10.1049/cp.2011.0758}
\showDOI{\tempurl}


\bibitem[\protect\citeauthoryear{Young and La~Joie}{Young and La~Joie}{2009}]%
        {metadataexchangeconcepts}
\bibfield{author}{\bibinfo{person}{Ian~A. Young} {and} \bibinfo{person}{Chad
  La~Joie}.} \bibinfo{year}{2009}\natexlab{}.
\newblock \bibinfo{title}{{Interfederation and Metadata Exchange: Concepts and
  Methods}}.
\newblock
  \bibinfo{howpublished}{\url{http://iay.org.uk/blog/2009/05/concepts-v1.10.pdf}}.
\newblock
\newblock
\shownote{[Online, \today].}


\bibitem[\protect\citeauthoryear{{Yuan Cao} and {Lin Yang}}{{Yuan Cao} and {Lin
  Yang}}{2010}]%
        {5689468}
\bibfield{author}{\bibinfo{person}{{Yuan Cao}} {and} \bibinfo{person}{{Lin
  Yang}}.} \bibinfo{year}{2010}\natexlab{}.
\newblock \showarticletitle{{A survey of Identity Management technology}}. In
  \bibinfo{booktitle}{\emph{{2010 IEEE International Conference on Information
  Theory and Information Security}}}. \bibinfo{pages}{287--293}.
\newblock
\showISSN{null}
\urldef\tempurl%
\url{https://doi.org/10.1109/ICITIS.2010.5689468}
\showDOI{\tempurl}


\bibitem[\protect\citeauthoryear{{Zefferer}, {Ziegler}, and
  {Reiter}}{{Zefferer} et~al\mbox{.}}{2017}]%
        {8356430}
\bibfield{author}{\bibinfo{person}{Thomas {Zefferer}}, \bibinfo{person}{Dominik
  {Ziegler}}, {and} \bibinfo{person}{Andreas {Reiter}}.}
  \bibinfo{year}{2017}\natexlab{}.
\newblock \showarticletitle{{Best of Two Worlds: Secure Cloud Federations meet
  eIDAS}}. In \bibinfo{booktitle}{\emph{2017 12th International Conference for
  Internet Technology and Secured Transactions (ICITST)}}.
  \bibinfo{pages}{396--401}.
\newblock
\urldef\tempurl%
\url{https://doi.org/10.23919/ICITST.2017.8356430}
\showDOI{\tempurl}


\bibitem[\protect\citeauthoryear{Zhang, Cho, Wu, and Shieh}{Zhang
  et~al\mbox{.}}{2015}]%
        {7185289}
\bibfield{author}{\bibinfo{person}{Zhi-Kai Zhang},
  \bibinfo{person}{Michael~Cheng~Yi Cho}, \bibinfo{person}{Zong-Yu Wu}, {and}
  \bibinfo{person}{Shiuhpyng~Winston Shieh}.} \bibinfo{year}{2015}\natexlab{}.
\newblock \showarticletitle{{Identifying and Authenticating IoT Objects in a
  Natural Context}}.
\newblock \bibinfo{journal}{\emph{Computer}} \bibinfo{volume}{48},
  \bibinfo{number}{8} (\bibinfo{date}{Aug} \bibinfo{year}{2015}),
  \bibinfo{pages}{81--83}.
\newblock
\showISSN{0018-9162}
\urldef\tempurl%
\url{https://doi.org/10.1109/MC.2015.213}
\showDOI{\tempurl}


\bibitem[\protect\citeauthoryear{{Zouari} and {Hamdi}}{{Zouari} and
  {Hamdi}}{2016}]%
        {7746120}
\bibfield{author}{\bibinfo{person}{Jaweher {Zouari}} {and}
  \bibinfo{person}{Mohamed {Hamdi}}.} \bibinfo{year}{2016}\natexlab{}.
\newblock \showarticletitle{{AIDF: An Identity as a Service Framework for the
  Cloud}}. In \bibinfo{booktitle}{\emph{2016 International Symposium on
  Networks, Computers and Communications (ISNCC)}}. \bibinfo{pages}{1--5}.
\newblock
\urldef\tempurl%
\url{https://doi.org/10.1109/ISNCC.2016.7746120}
\showDOI{\tempurl}


\end{thebibliography}

\end{document}